\documentclass[sn-mathphys-num]{sn-jnl}
\usepackage{amsmath,amssymb,amsfonts}
\usepackage{amsthm}
\usepackage{mathrsfs}
\usepackage[title]{appendix}
\usepackage{xcolor}
\usepackage{textcomp}
\usepackage{manyfoot}
\usepackage{booktabs}
\usepackage{algorithm}
\usepackage{algorithmicx}
\usepackage{algpseudocode}
\usepackage{listings}
\usepackage{gensymb}
\usepackage{array}
\usepackage{adjustbox}
\usepackage{natbib}
\usepackage{multirow}
\usepackage{subfigure}
\setcitestyle{authoryear,open={(},close={)}}

\theoremstyle{thmstyleone}

\theoremstyle{thmstyletwo}

\theoremstyle{thmstylethree}

\begin{document}

\title[Article Title]{Towards High Resolution Real-Time Optical Flow Particle Image Velocimetry}

\author*[1,2]{\fnm{Juan} \sur{Pimienta}}\email{juan.pimienta@espci.fr}

\author*[1]{\fnm{Jean-Luc} \sur{Aider}}\email{jean-luc.aider@espci.fr}

\affil*[1]{\orgdiv{Laboratoire PMMH}, \orgname{ESPCI Paris - PSL, CNRS}, \orgaddress{\street{7-9 quai Saint Bernard}, \city{Paris}, \postcode{75005}, \country{France}}}

\affil[2]{\orgdiv{Photon Lines}, \orgaddress{\street{10 avenue des Touches}, \city{Pacé}, \postcode{35740},  \country{France}}}

\abstract{Particle Image Velocimetry (PIV) is the most commonly used optical technique for measuring 2D velocity fields. However, improving the spatial resolution of instantaneous velocity fields and having access to the velocity field in real time remains a challenge. Optical Flow veolcimetry makes it possible to meet these challenges. In this study, we show that it is possible to access dense velocity fields (1 vector per pixel) in real-time using an appropriate seeding concentration adapted to optical flow algorithms and no longer to cross-correlation PIV algorithms. The influence of concentration on the quality of velocity fields is demonstrated using synthetic images generated for a Rankine vortex. We thus demonstrate that it is possible to precisely measure small vortices using optical flow provided that the seeding is suitable. The notion of "Active Pixels" is also introduced in order to define a seeding optimization criterion adapted to experimental measurements. This criterion is finally successfully applied to the flow downstream of a cylinder leading to a spatial resolution down to one vector per pixel.}

\keywords{Optical Flow, Particle Image Velocimetry, Seeding, Concentration of Particles, High Resolution, Real-Time measurements. }

\maketitle

\section{Introduction}
Particle Image Velocimetry (PIV) is a non-intrusive optical technique that allows the measurement of the two-components (2C) of a velocity fields in a plane (2D) defined by a laser light sheet traversing a fluid flow seeded with reflecting particles \citep{raffel2018particle}. The basic principle consists in computing the displacement of the particles between two successive snapshots using, in standard PIV post-processing, a FFT (Fast Fourier Transform) cross-correlation (CC) algorithm. CC-PIV is the standard algorithm currently used in most experiments, despite being very time consuming, computationally demanding and limited in terms of spatial resolution or real-time measurements.  To optimize the quality of the velocity fields, it is important to choose the proper experimental parameter adapted to the CC algorithms, like the time interval between two snapshots, related to a maximum displacement of a few particles inside an Interrogation Windows (IWs), which is a key element for the spatial resolution of the PIV field \citep{kahler2012resolution}. 

However, the optimal parameters for other type of algorithms, like Optical Flow (OF), may not be the same as the ones for a CC algorithm. Indeed, OF-PIV offers a different approach to estimate the velocity fields through the motion of particles. Coming from the community of Machine Vision, Optical Flow can be understood as the apparent velocities from changes in intensity patterns in a scene \citep{gibson_perception}. From this point of view, the particles are only used as a way to create a textured image that is modified by the flow between two time steps. The general idea to estimate the displacements from intensity changes is then based on the assumption that intensity levels are kept constant between successive frames and that displacements are supposed to be small, of the order of 1 pixel.

Determining displacement vectors from intensity variations is an under-constrained problem. This problem was solved mainly in two ways. Either a smoothness constraint is imposed on the system (Horn-Schunck algorithm) \citep{HORN1981185} which produces a global solution, or it is assumed that the displacements in the vicinity of a kernel centered on a pixel are very close to each other (Lukas -Kanade algorithm) \citep{Lucas1981AnII}. Later, the Lukas-Kanade OF algorithm was modified by adding an iterative scheme (Folki) \citep{Besnerais2005DenseOF} and then adapted to perform PIV calculations \citep{Champagnat2011FastAA}. One of the considerable advantages of OF-PIV algorithms is that they can be easily parallelized, particularly on GPUs (Graphics Processor Unit) whose architecture makes them an ideal partner for OF algorithms. This is why the Folki algorithm has been optimized to operate in real time \citep{Gautier2013RealtimePF} to the point of being used as a sensor in closed-loop flow control experiments \citep{Gautier2013ControlOT,Gautier2015FrequencylockRC}. In addition to the significant gain in  calculation time \citep{Plyer2016MassivelyPL}, OF-PIV algorithms should also lead to \textit{dense} velocity fields, with a spatial resolution which should reach 1 vector per pixel. Although OF-PIV has indeed been shown to lead to smaller scales than CC-PIV in turbulent spectra \citep{giannopoulos2022optimal}, the resolution of one vector per pixel has not been achieved.

In recent years, the entire acquisition chain for real-time PIV measurement has been optimized (Laser, high-speed streaming cameras, dedicated computer, optimization of the algorithm). It leads to a high-resolution, high-frequency real-time optical flow PIV (RT-OFPIV) system. Being able to run RT-OFPIV and calculate quantities derived from the instantaneous velocity field in real time has made possible almost unlimited observations, analysis or recording of a flow \citep{pimienta2022characterization} leading to the possibility to run machine learning or neural network analysis based on large OF-PIV databases \citep{gautier2015closed,giannopoulos2020data}. This also leads to new experimental challenges. For example, various studies and optimizations have been carried out to improve the experimental conditions, the selection of algorithm parameters or the selection of appropriate equipment.

The present study is divided into two main sections. The first part consists of a benchmark of OF-PIV based on synthetic particles images generated by a given flow (Rankine vortex).  The influence of various image parameters (particle concentration, displacement amplitude) on the spatial resolution and velocity estimation at various scales is studied. The second part is focused on the application of the optimization of the experimental parameters in order to improve the quality the RT-OFPIV experimental measurements on two test cases: an uniform vortex-free flow and a massively separated flow downstream a cylinder.

\section{Experimental set up}

\subsection{Hydrodynamic channel}

\begin{figure}[!h]
    \centering
    \subfigure(a){\includegraphics[width=0.8\textwidth]{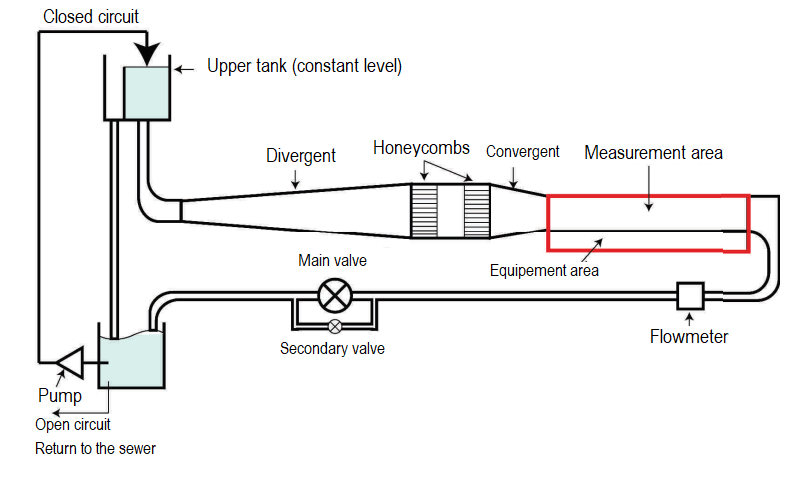}}\\
    \subfigure(b){\includegraphics[width=0.6\textwidth]{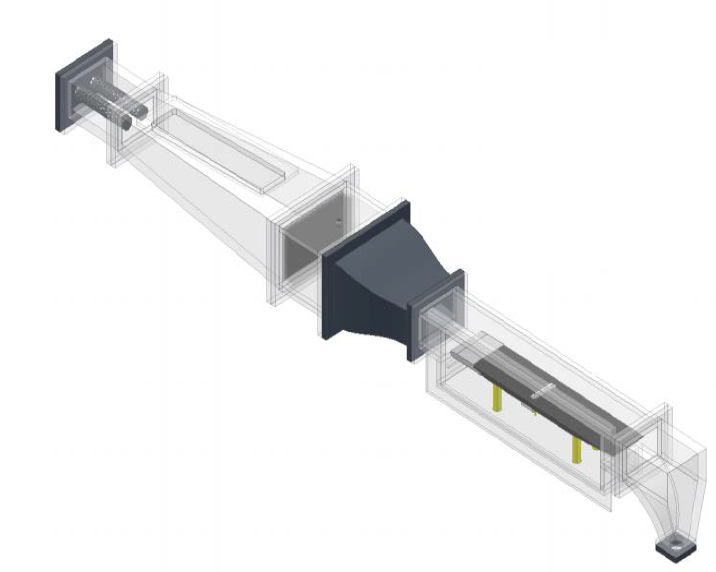}}
    
    \caption{a) Sketch of the hydrodynamic channel. The flow is driven by gravity and stabilized using honeycombs upstream of the test section \citep{T.cambonie:pastel-00795587}. b) 3D Sketch of test section which contains a flat plate allowing the growth of a boundary layer upstream of a the Cylinder \citep{gautier:tel-01150428}.}
    \label{fig:Channel}
\end{figure}

Experiments have been carried out in a hydrodynamic channel in which the flow is driven by gravity (Fig.~\ref{fig:Channel}a), using a constant level water tank to ensure a pressure differential of $\Delta P = 0.3$~$bar$. The maximum free-stream velocity $U_{\infty}=22$~$cm.s^{-1}$. The flow is stabilized by divergent and convergent sections separated by honeycombs, leading to a turbulence intensity lower than  $1$~\%. A NACA 0020 profile is used to smoothly start a Blasius boundary layer over the flat plate, upstream of the separated flow. The test section is $80~cm$ long with a rectangular cross-section $w=15$~$cm$  wide and $H=7.7~cm$ high (Fig.~\ref{fig:Channel}b). 

To study a wake flow, a vertical cylinder is mounted over the flat plate. It has a diameter of $D=1$~$cm$ and is placed at $x=40$~$cm$ from the leading edge of the plate. The maximum Reynolds number based on the diameter of the cylinder $D$ is $Re_D = \frac{U_\infty * D}{\nu} \approx 2200$ for a water temperature of $21^{\circ}$C ($\nu$ being the kinematic viscosity of water).

\subsection{OF-PIV setup}

\begin{figure}[!h]
    \centering
    \includegraphics[width=0.8\textwidth]{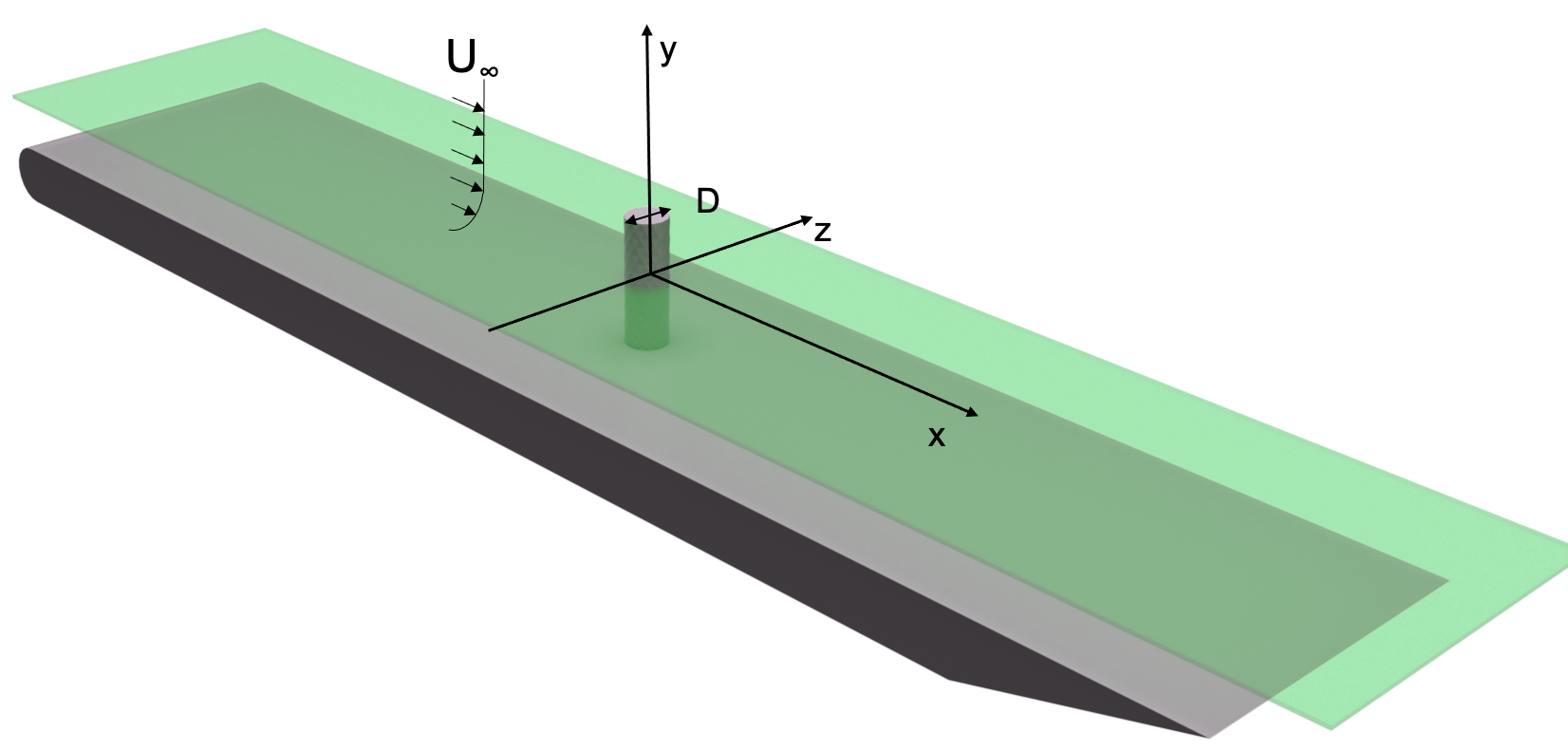}
    \caption{3D Sketch of the flat plate and the cylinder model. The profiled horizontal plate, allowing the development of a boundary layer upstream of the cylinder, is installed inside the rectangular test section shown in Fig.~\ref{fig:Channel}b. The optical flow measurements were carried out in a horizontal plane at mid-height (green plane here) of the cylinder, both in the free-stream region, upstream of the cylinder, and in the wake of the cylinder.}
    \label{fig:Cyl}
\end{figure}

To carry out the PIV measurements, the water was seeded with light reflecting polyamid neutrally buoyant micro particles of $20~\mu m$ of diameter. The flow was illuminated by a laser sheet generated by a laser beam going through a Powell cylindrical lens. A Coherent\texttrademark continuous Nd:Yag laser (wavelength of 532~$nm$ for a power output of 2~$W$) was used. The measurements were carried out in an horizontal plane at the mid-height of the channel, covering the free-stream region upstream of the cylinder as well as the region downstream of a cylinder (Fig.~\ref{fig:Cyl}). 

To record the instantaneous snapshots of the seeded flow, a Mikrotron\texttrademark 21CXP12 camera was used. It allows the acquisition of 21~$Mpx$ images with an acquisition frequency up to 240~$Hz$, which can be streamed toward a dedicated workstation through a CoaxPress card. 

A computer has been designed and built to optimize its performances for real-time acquisition. The system is based on a AMD Ryzen Threadripper PRO 3955WX  processor with 16 cores running at a frequency of 3.90~GHz with 128~GB of RAM. Two powerful, last generation, GPUs (RTX4070) are supported on a custom open chassis that allows for better access and easier connection/removal of new GPUs.

\section{Real-Time Optical Flow PIV}

The optical flow algorithm is based upon the assumption of intensity conservation between successive images, which can be expressed as: 

\begin{equation}
    \nabla I(x,y,t) = 0
    \label{eq:intensity_gradient}
\end{equation}

where $I(x,y,t)$ is the light intensity at every pixels of the images at a given time $t$.

Using this assumption (Eq.~\ref{eq:intensity_gradient}) and an additional constraining condition, it is theoretically possible to estimate the flow displacement between two successive snapshots at each pixel contained in the images. The constraint imposed by the Lukas-Kanade \citep{Lucas1981AnII} method is to assume that neighboring pixels will behave in a similar way. This is the reason why FOLKI \citep{Besnerais2005DenseOF} can be considered as a compromise between pure optical flow methods and window-based methods. This is because it uses a pixel-centered \textit{kernel}, which defines one of the important parameters of the algorithm called the \textit{kernel radius} (KR). It defines the size of the areas where the intensity gradients will be compared. This process will be applied to each pixel, leading to a resolution of one vector per pixel. This is very different from the interrogation windows (IW) used in the CC-PIV standard, which defines a minimum area containing a few particles that will move inside the IW between two time steps.

Optical flow codes can be limited to estimating small displacements, of the order of 1 pixel. However, this problem is solved by the implementation of a Gaussian pyramid scheme, which allows successive reductions in the size of the image, therefore subsampling of large displacements. This defines another important parameter of the algorithm called the \textit{pyramid sublevels}. At each new pyramid sublevel, the number of pixels in each direction is halved.

The last important parameter is the number of Gauss-Newton iterations that the code must perform to give a solution. An extra step of pre-processing has been added before calculating the velocity fields, which consists in an intensity normalization using a pixel-centered kernel all across the image.

The whole process of velocity fields calculation consists in six main steps:

\begin{enumerate}
    \item  Normalization of intensity of the image.
    \item  Image sub-sampling with Gaussian pyramids.
    \item  Estimation of the displacements at the kernel scale.
    \item  Projection of the velocity fields up-sampling the image size.
    \item  Iterations through the user defined times.
    \item  Velocity fields estimation.
\end{enumerate}

\begin{figure}[!h]
    \centering
    \includegraphics[width=0.8\textwidth]{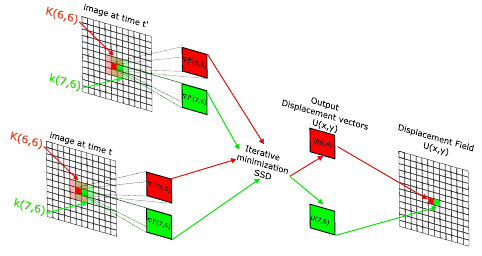}
    \caption{Diagram showing the main steps used to compute the velocity vectors at the kernel scale.}
    \label{fig:KR}
\end{figure}

The process of estimating displacements, particularly at the kernel scale, is illustrated in Fig.~\ref{fig:KR}. This starts with two subsamples of size $12~\times~12$ pixels, for two successive times $t$ and $t' = t+\delta t$. In this example, two neighboring pixels at positions $(x, y) = (6, 6)$ (red pixel) and $(x, y) = (7, 6)$ (green pixel) are considered with a  Kernel radius $KR = 1$ pixel. The core associated with each pixel can be seen as a lighter shade of the pixel's color (left side of Fig~\ref{fig:KR}). It is at this scale that the spatial intensity gradients are calculated, then the difference minimized using the iterative Gauss-Newton scheme to find the displacement leading to the velocity vectors for each of the positions evaluated. This process is performed for each pixel of successive snapshots, yielding a displacement vector per pixel.  In the following, the tests will be carried out with EyePIV\texttrademark, a dedicated OF-PIV plugin developed in collaboration between the two teams (Laboratory PMMH in collaboration with Photon Lines) for real-time measurements.   \\

\section{Synthetic images generation}

To assess the quality of the results delivered by the OFPIV, a benchmark with synthetic PIV images was carried out. Synthetic PIV images were generated using a \textit{PIV-image-generator} developed by \citep{MENDES_piv_image_gen}. This Matlab-based software allows the synthetic creation of PIV images, using a well-known analytical expression of specific flows, such as a shear flow or a Rankine vortex, which will be used to move a set of randomly distributed particles in a given snapshot.

In the following, the accuracy and precision of OF-PIV measurements will be studied using different image parameters and for a given flow: a single Rankine vortex. The Rankine vortex is defined using the analytical expressions summarized in Tab.~\ref{tab:flows} as proposed by \citep{MENDES_piv_image_gen}. This is a very interesting benchmark due to the spatially localized high velocity gradients present in this type of flow, as well as the possibility of imposing important and critical parameters for PIV measurements, such as the vortex radius and the amplitude of the rotation speed. These features were identified as challenging for CC-PIV \citep{Beresh_velocity_gradient_PIV_turbulence, westerweel_piv_gradients}.   

\begin{table}[!ht]
    \centering
    \begin{tabular}{|c|p{6cm}|p{4cm}|}
        
        \hline
        \centering
        \textbf{Flow type} & \textbf{Displacement in the flow field} & \textbf{Velocity flow field}  \tabularnewline
        \hline
        \centering

         Rankine vortex & \parbox{4cm}{\begin{align*}\begin{array}{lr}x_1 &= r\cos{\bigl(\frac{\Gamma}{2\pi R}\frac{1}{MR}t+\theta \bigr)}\\ ~~\\ y_1 &= r\sin{\bigl(\frac{\Gamma}{2\pi R}\frac{1}{MR}t+\theta \bigr)}\end{array}\end{align*}} , $r\leq R$ 
         \newline \parbox{4cm}{\begin{align*}\begin{array}{lr}x_1 &= r\cos{\bigl(\frac{\Gamma}{2\pi R}\frac{1}{Mr^2}t+\theta \bigr)}\\ ~~\\ y_1 &= r\sin{\bigl(\frac{\Gamma}{2\pi R}\frac{1}{Mr^2}t+\theta \bigr)}\end{array}\end{align*}} , $r > R$ & \begin{equation*}u_\theta(r) = \begin{cases} \frac{\Gamma}{2\pi R}\frac{r}{MR} , r\leq R\\~~\\ \frac{\Gamma}{2\pi R}\frac{r}{Mr}, r>R \end{cases}\end{equation*} \tabularnewline
         \hline

    \end{tabular}
    \caption{Flow type description: $m$ is a real displacement, $M$ is a size factor, $\Gamma $ is the circulation of the vortex, $\theta$ the azimuthal polar coordinate, $r$ is the radial polar coordinate, $t$ is the time.}
    \label{tab:flows}
\end{table}

The generation of synthetic PIV images requires the definition of a set of parameters. Some parameters remained constant for generating all PIV images, such as image size ($1024~\times~1024~pixels^2$), particle radius ($r_p = 1.5~pixels$ ), the radius of the simulated particles, the thickness of the laser sheet ($0.2~mm$), the standard deviation of the out-of-plane motion ($\sigma = 0.025$) and no added noise. The parameters studied were the maximum expected displacement, the particle concentration, expressed in particles per interrogation window [part/IW] and the core radius of the Rankine vortex ($R$). The IW size used to generate the images was $IW~=16~px$. This definition of particle concentration is convenient for CC-PIV but not really well suited for OF-PIV and will be discussed in more detail in section ~\ref{sec:exp_con}. Table~\ref{tab:images} lists the dynamic parameters used to create the images. For reasons of statistical relevance, ten pairs of images were created for each of the possible combinations of dynamic parameters. As a result, there were 216 sets of 10 even images to perform the benchmark.

\begin{table}[!h]
    \centering
    \begin{tabular}{|p{4cm}|p{4cm}|p{4cm}|}
    \hline
     \textbf{Maximum \newline displacement [pixels]} & \textbf{Particles concentration \newline [particles / IW]} & \textbf{Rankine core \newline radius [pixels]}\\
     \hline
     8, 16, 24, 32 & 5, 10, 15 & 12, 25, 50, 75, 100, 150\\
     \hline
    \end{tabular}
    \caption{Parameters used for the generation of the synthetic images. \textit{IW} stands for interrogation window.}
    \label{tab:images}
\end{table}

The OF-PIV parameter space has been kept reasonably constrained since it has been empirically found that there is a "convergence zone", meaning that any increment from this region will only lead to a deterioration of the results, particularly when dealing with high gradient flows. The OF-PIV parameters are reported in Table ~\ref{tab:OFPIV} as minimum, maximum and per increment, leading to 288 unique combinations. In general, for each image parameter spanning the OFPIV parameter combination, there were 2 880 data entry points. This represents 207 360 data entry points from 414 720 image pairs for the entire study.

\begin{table}[!h]
    \centering
    \begin{tabular}{|c|c|c|c|}
        \hline
         \textbf{Parameter} & \textbf{Minimum}  & \textbf{Maximum} & \textbf{Step}\\
         \hline
         \textbf{Kernel Radius} & 2 & 7 & 1 \\
         \hline
         \textbf{Pyramid sub-level} & 2 & 4 & 1 \\
         \hline
         \textbf{Iterations} & 1 & 4 & 1 \\
         \hline
         \textbf{Normalization Radius} & 1 & 4 & 1 \\
         \hline
    \end{tabular}
    \caption{Main parameters of the OF-PIV with their ranges of variation used in this study.}
    \label{tab:OFPIV}
\end{table}

\begin{figure}[!h]
    \centering
    \includegraphics[width = 1\textwidth]{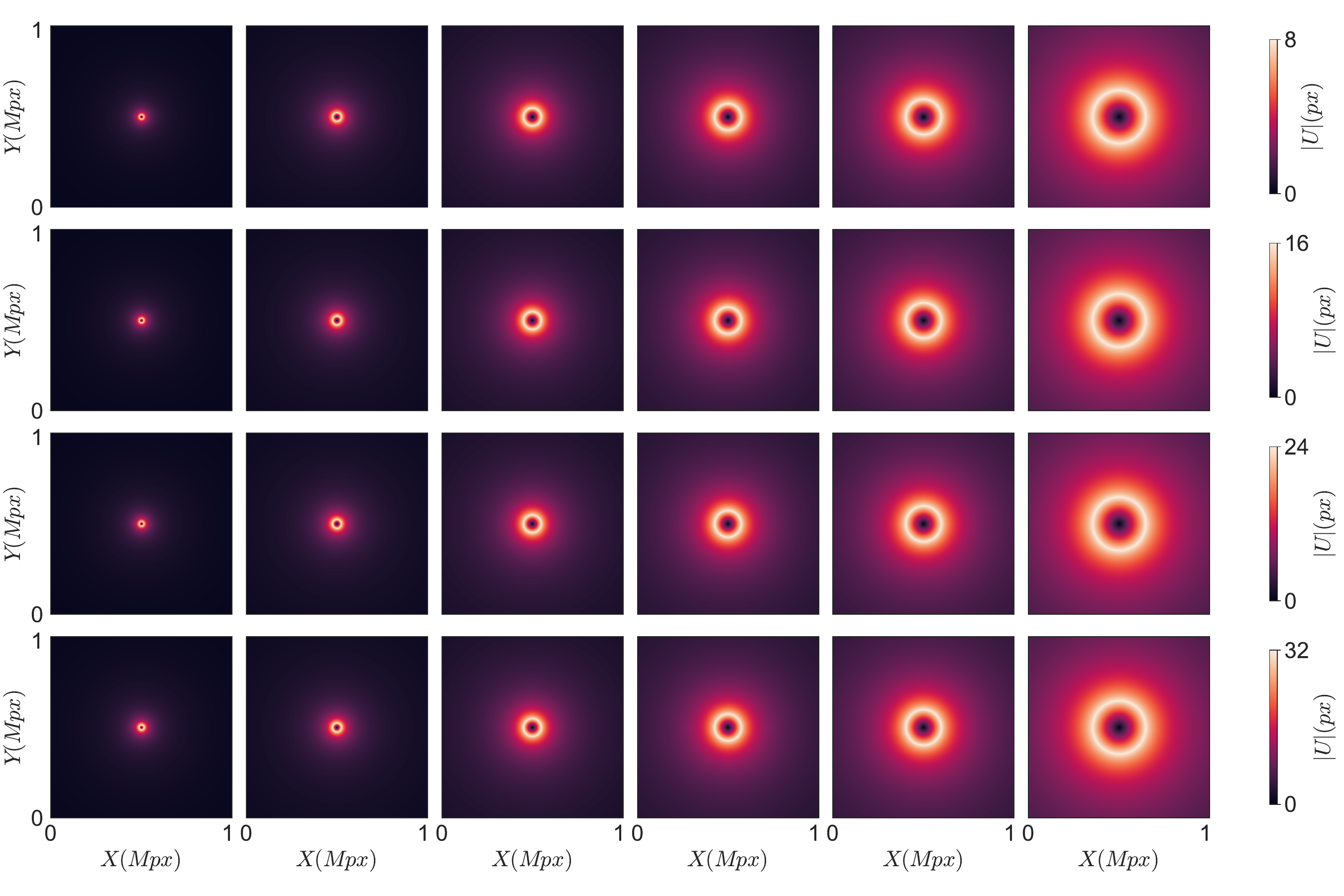}
    \caption{Theoretical displacement fields of the Rankine vortex for various radius and velocities. The core radius of the vortex changes along the X axis (from $r = 12$ to $r=150$ pixels) while the maximum displacement of the field changes along the Y axis (from 8 to 32 pixels from top to bottom).}
    \label{fig:th_displacement}
\end{figure}

The velocity fields corresponding to various synthetic Rankine vortices obtained for different parameters are shown in Fig.~\ref{fig:th_displacement}. The central radius of the Rankine vortex increases from left to right, while the maximum magnitude of displacement increases from top to bottom. As the size of the images remains constant, this means that OF-PIV will be tested on large vortices as well as very small vortices, for low and high rotations, which are the key characteristics needed to correctly measure, for example, turbulent shear flows.

\section{Influence of concentration on the quality and spatial resolution of OF-PIV}

\subsection{Error estimation}

\begin{figure}[!h]
    \centering
    \includegraphics[width=1\textwidth]{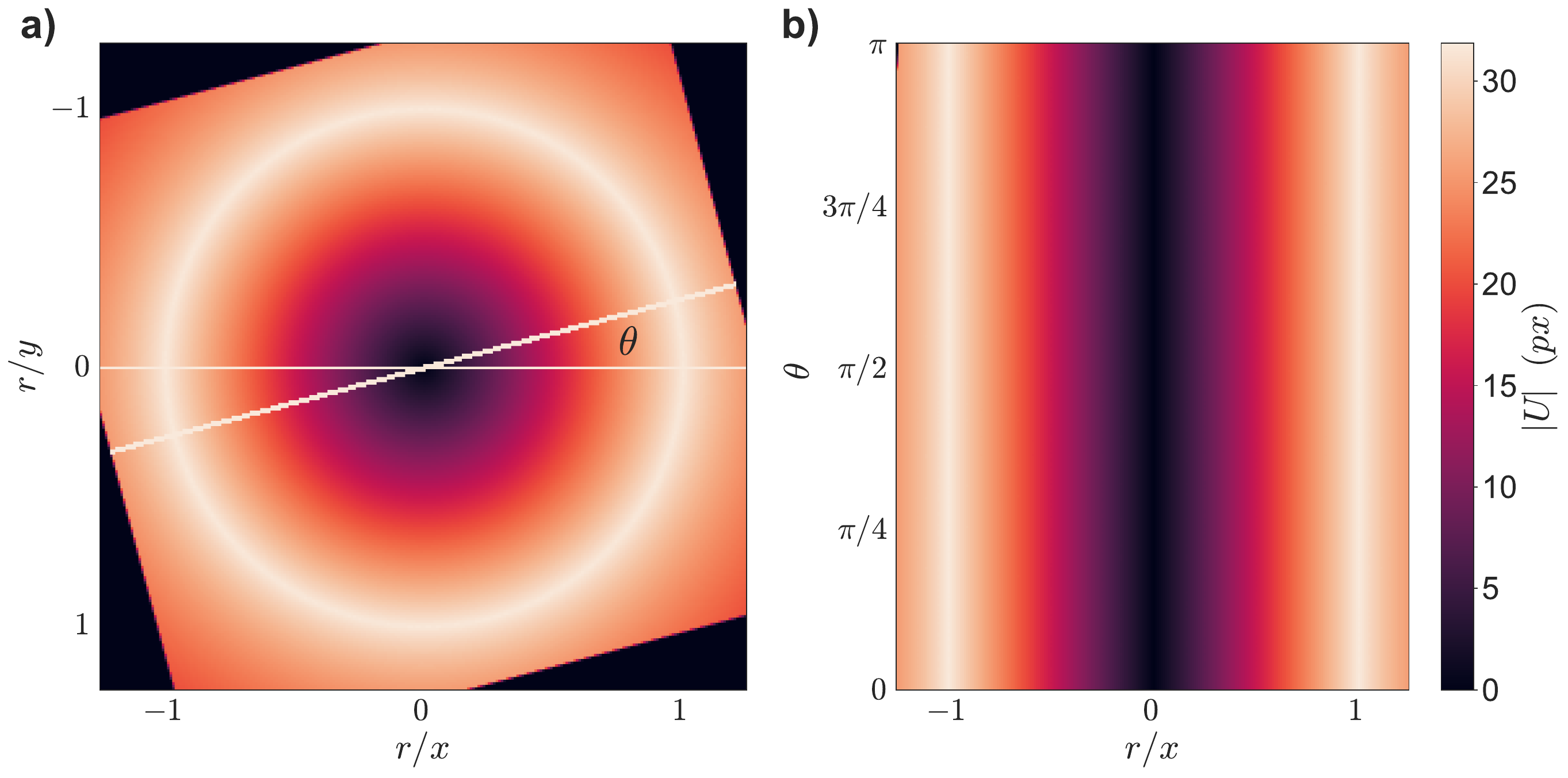}

    \caption{a) {Illustration of the principle used to obtain all the displacement profiles defining the Rankine vortex through a rotation of the velocity field. b) Contour plot of all the velocity profiles defining the Rankine vortex as a function of the rotation angle $\theta$. Example with a Rankine vortex with a core radius of 100 pixels and a maximum displacement of 32 pixels.}}
    \label{fig:rankine_stretch}
\end{figure}

Two error estimation criteria were used for this study to quantify the quality of the OF-PIV calculations: the absolute displacement error and a comparison of the displacement profile of the Rankine vortex. The absolute displacement error $Err$ is defined as:

\begin{align}
    Err(x,y) = \sqrt{(u_{th} - u_{OF})^2 + (v_{th} - v_{OF})^2}
    \label{eq:absolute_error}
\end{align}

where subscripts \textit{th} and \textit{OF} stands respectively for the theoretical displacement and the OF-PIV computed result. $u$ and $v$ are respectively the velocities along the $x$ and $y$ directions.

Comparison of displacement profiles was performed by extracting all velocity profiles across the vortex core. This process first involves subsampling the image to better target the vortex in the center of the image. Second, a one-pixel velocity profile $U(x,y) = \sqrt{u(x,y)^2 + v(x,y)^2}$ is selected in the middle of the image ($y/2$) and is stored in an array $R$. Third, the downsampled image is rotated by a specific angle $\theta$ ($rot = f(U(x,y),\theta)$) to achieve a rotation equivalent to 1 pixel (Fig.~\ref{fig:rankine_stretch}a). The rotation angle changes depending on the radius of the vortex core. Fourth, a new row of one pixel at $y/2$ is selected in the rotated field and stored in the $R(x, y)$ array. Steps 3 and 4 are repeated until a rotation of 180 $\degree$ is obtained. Fig.~\ref{fig:rankine_stretch}b shows an example of all the velocity profiles plotted against angle $\theta$ for a theoretical Rankine vortex, which is why the same profile is obtained for each angle $\theta$.

\begin{figure}[!h]
    \centering
    \includegraphics[width = 0.6\textwidth]{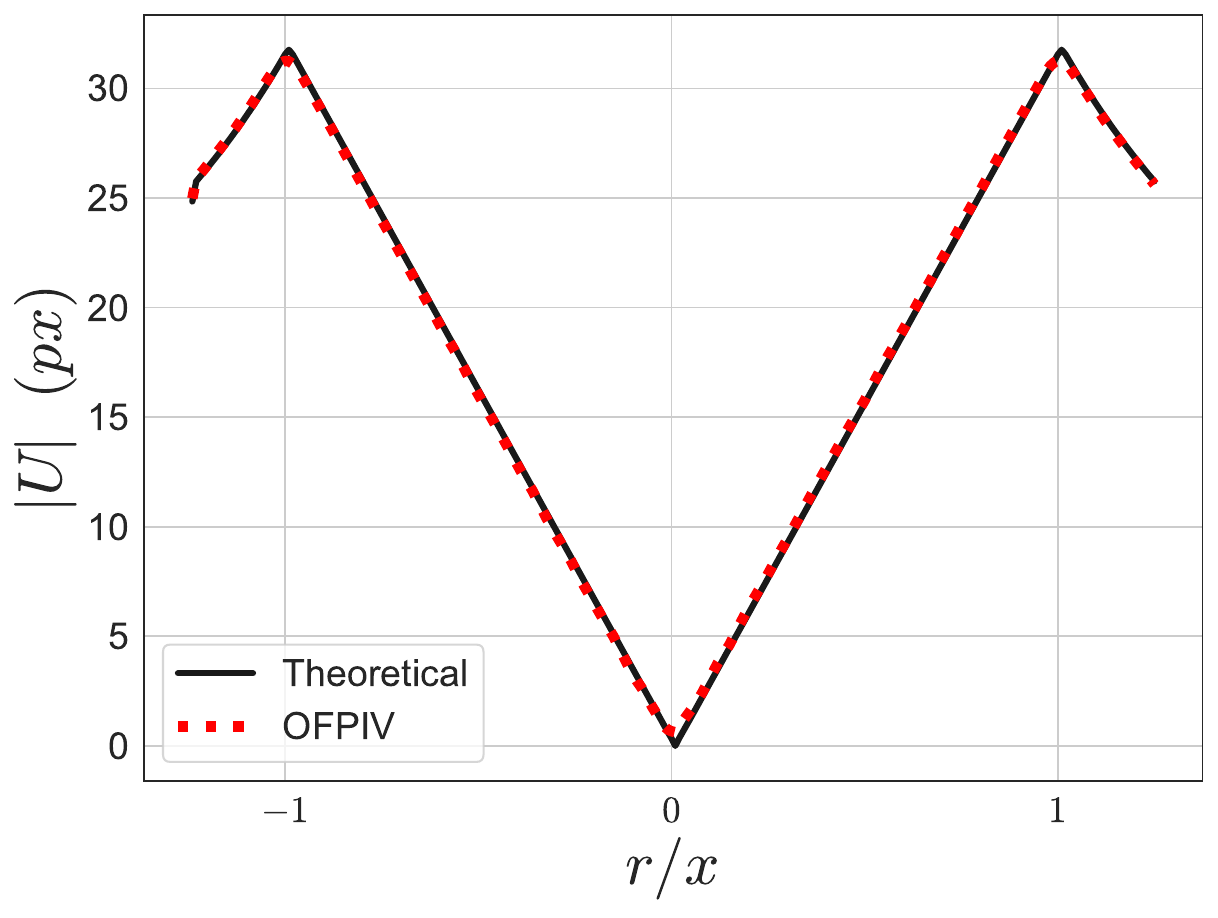}
    \caption{Comparison of the theoretical and OF-PIV displacement profiles obtained for a Rankine vortex. The radial profile is the result of the spatial average along the $\theta$ direction, as illustrated in Fig.~\ref{fig:rankine_stretch}b. This result is obtained for a Rankine vortex with a radius of 100 pixels and a maximum displacement of 32 pixels.}
    \label{fig:profile_comparison}
\end{figure}

This process is carried out both for the theoretical domains and for the OF-PIV domains. Then the two fields are compared. A criterion has been defined to store the error values as a scalar value. A criterion $\Psi$ is defined as the sum of the absolute differences between the theoretical displacement field \textit{$\theta$ averaged} and the results \textit{unfolded} OF-PIV: $\Psi = \Sigma^{i,j=x,y}_{i,j=0} |R_{th}(i,j) - R_{ofpiv}(i,j)|$. For clarity, the criteria are normalized with the vortex area \textit{unfolded}.

\begin{figure}[!h]
    \centering
    \includegraphics[width=1\textwidth]{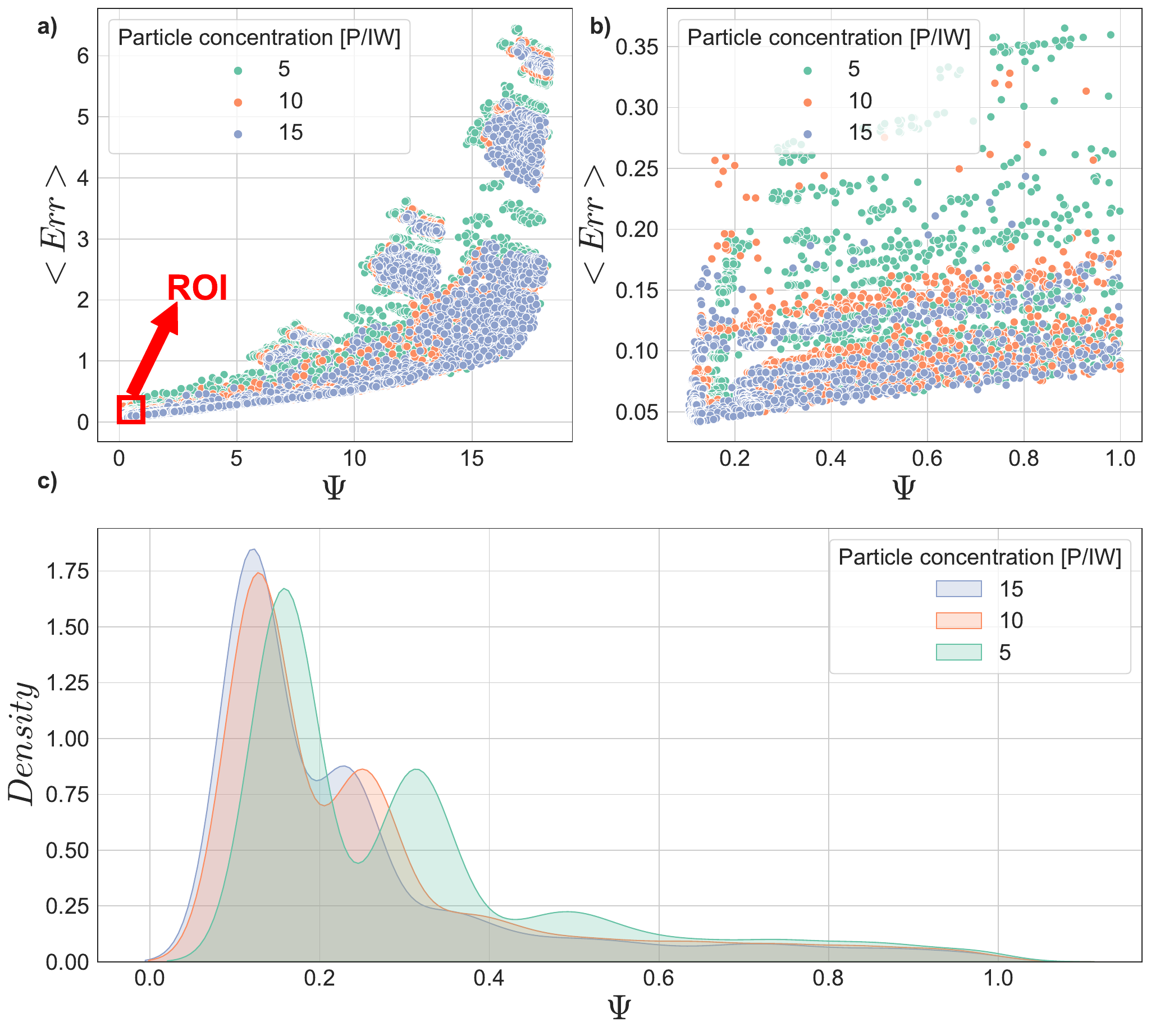}
    \caption{Scatter plot of error analysis for a Rankine vortex with radius of 75 pixels. \textbf{a)} Scatter plot of the both error metrics for all displacements and concentrations. \textbf{b)} Scatter plot of the ROI. \textbf{c)} Probability density distribution of $\Psi$ on the ROI.}
    \label{fig:rankine_core_r75}
\end{figure}

Fig.~\ref{fig:profile_comparison} shows the comparison between the theoretical and OF-PIV spatially averaged (along the $\theta$ direction) displacement profiles. In this particular example a very good agreement can be observed, both for the amplitude and location of the maxima of velocity. It will be further detailed and discussed in the Results section.

\subsection{Influence of particle concentration on the quality of OF-PIV computations}

Fig.~\ref{fig:rankine_core_r75} shows a scatter plot summarizing the error analysis as a function of the different particle concentrations obtained for the case of the Rankine Vortex with a vortex core of $r = 75 ~~pixels $ and maximum displacements. It is clearly seen that, for a given set of OF-PIV parameters, the higher the concentration of particles in the images, the lower the error levels. 

More precisely, Fig.~\ref{fig:rankine_core_r75}a) is a scatter plot of the spatially averaged displacement error over the entire image $<Err>$ and \textit{unfolded} profile criteria  $\Psi$ through the different maximum expected displacements and particle concentrations. Fig.~\ref{fig:rankine_core_r75} b) is a zoom in on a specific region of interest (ROI) in Fig~\ref{fig:rankine_core_r75}a) where the results show minimal error with both criteria. It corresponds to $\Psi$ values less than 1. The criterion being normalized by the area of the vortex \textit{unfolded}, this value can be interpreted as an average difference between the theoretical displacements and OF-PIV of 1 pixel.
Fig~\ref{fig:rankine_core_r75}c) shows the probability density function (PDF) of the $\Psi$ criteria inside the ROI for the three concentrations. Here it can be clearly understood that for all different displacement magnitudes, concentration plays a major role in the error margins.

\begin{table}[!ht]
    \centering
    \begin{adjustbox}{width=1\textwidth}
        \begin{tabular}{|>{\centering\arraybackslash}m{.15\linewidth}|>{\centering\arraybackslash}m{.15\linewidth}|>{\centering\arraybackslash}m{.2\linewidth}|>{\centering\arraybackslash}m{.1\linewidth}|>{\centering\arraybackslash}m{.1\linewidth}|
        >{\centering\arraybackslash}m{.1\linewidth}|
        >{\centering\arraybackslash}m{.1\linewidth}|
        >{\centering\arraybackslash}m{.1\linewidth}|}
        \hline
             \textbf{Rankine \newline core radius\newline (px)}&\textbf{Maximum\newline displacement (px)}&\textbf{Particle\newline concentration (p/IW)} & \textbf{KR\newline(px)}&\textbf{NR\newline(px)}&\textbf{PSL}&\textbf{IT}&\textbf{D/R} \\
             \hline
             \multirow{4}{*}{12}& \multicolumn{1}{c|}{8}&\multicolumn{1}{c|}{15}&\multicolumn{1}{c|}{2}&\multicolumn{1}{c|}{1}&\multicolumn{1}{c|}{2}&\multicolumn{1}{c|}{4}&\multicolumn{1}{c|}{0.67}\\ 
            &16&5&2&3&4&4&1.33\\
            &24&15&2&2&4&4&2.00\\
            &32&10&2&1&4&4&2.67\\
             \hline
             \multirow{4}{*}{25}& \multicolumn{1}{c|}{8}&\multicolumn{1}{c|}{15}&\multicolumn{1}{c|}{3}&\multicolumn{1}{c|}{1}&\multicolumn{1}{c|}{3}&\multicolumn{1}{c|}{4}&\multicolumn{1}{c|}{0.32}\\
             &16&15&2&2&4&4&0.64\\
             &24&5&2&1&4&3&0.96\\
             &32&10&2&1&4&3&1.28\\
             \hline
             \multirow{4}{*}{50}& \multicolumn{1}{c|}{8}&\multicolumn{1}{c|}{15}&\multicolumn{1}{c|}{5}&\multicolumn{1}{c|}{1}&\multicolumn{1}{c|}{4}&\multicolumn{1}{c|}{3}&\multicolumn{1}{c|}{0.16}\\
             &16&15&4&1&3&4&0.32\\
             &24&10&2&3&4&4&0.48\\
             &32&5&3&1&4&4&0.64\\
             \hline
             \multirow{4}{*}{75}& \multicolumn{1}{c|}{8}&\multicolumn{1}{c|}{15}&\multicolumn{1}{c|}{4}&\multicolumn{1}{c|}{1}&\multicolumn{1}{c|}{4}&\multicolumn{1}{c|}{2}&\multicolumn{1}{c|}{0.11}\\
             &16&15&4&1&4&4&0.21\\
             &24&15&3&1&4&4&0.32\\
             &32&15&2&2&4&4&0.43\\
             \hline
             \multirow{4}{*}{100}& \multicolumn{1}{c|}{8}&\multicolumn{1}{c|}{15}&\multicolumn{1}{c|}{6}&\multicolumn{1}{c|}{1}&\multicolumn{1}{c|}{3}&\multicolumn{1}{c|}{4}&\multicolumn{1}{c|}{0.08}\\
             &16&15&4&1&3&4&0.16\\
             &24&15&3&1&4&4&0.24\\
             &32&15&2&1&4&4&0.32\\
             \hline
             \multirow{4}{*}{150}& \multicolumn{1}{c|}{8}&\multicolumn{1}{c|}{15}&\multicolumn{1}{c|}{6}&\multicolumn{1}{c|}{1}&\multicolumn{1}{c|}{4}&\multicolumn{1}{c|}{2}&\multicolumn{1}{c|}{0.05}\\
             &16&15&5&1&4&3&0.11\\
             &24&15&4&1&4&3&0.16\\
             &32&15&4&1&4&4&0.21\\
             \hline
        \end{tabular}
    \end{adjustbox}
    \caption{OF-PIV parameters configuration leading to the best possible result for each of the evaluated cases. \textbf{KR} stands for \textit{Kernel Radius}, \textbf{NR} for \textit{Normalization Radius}, \textbf{PSL} for \textit{Pyramid sub-level}, \textbf{IT} for \textit{Iterations} and \textbf{D/R} for the ratio between the maximum displacement and the Rankine core radius.}
    \label{tab:of_best}
\end{table}

\begin{figure}[!h]
    \centering
    \includegraphics[width=1\textwidth]{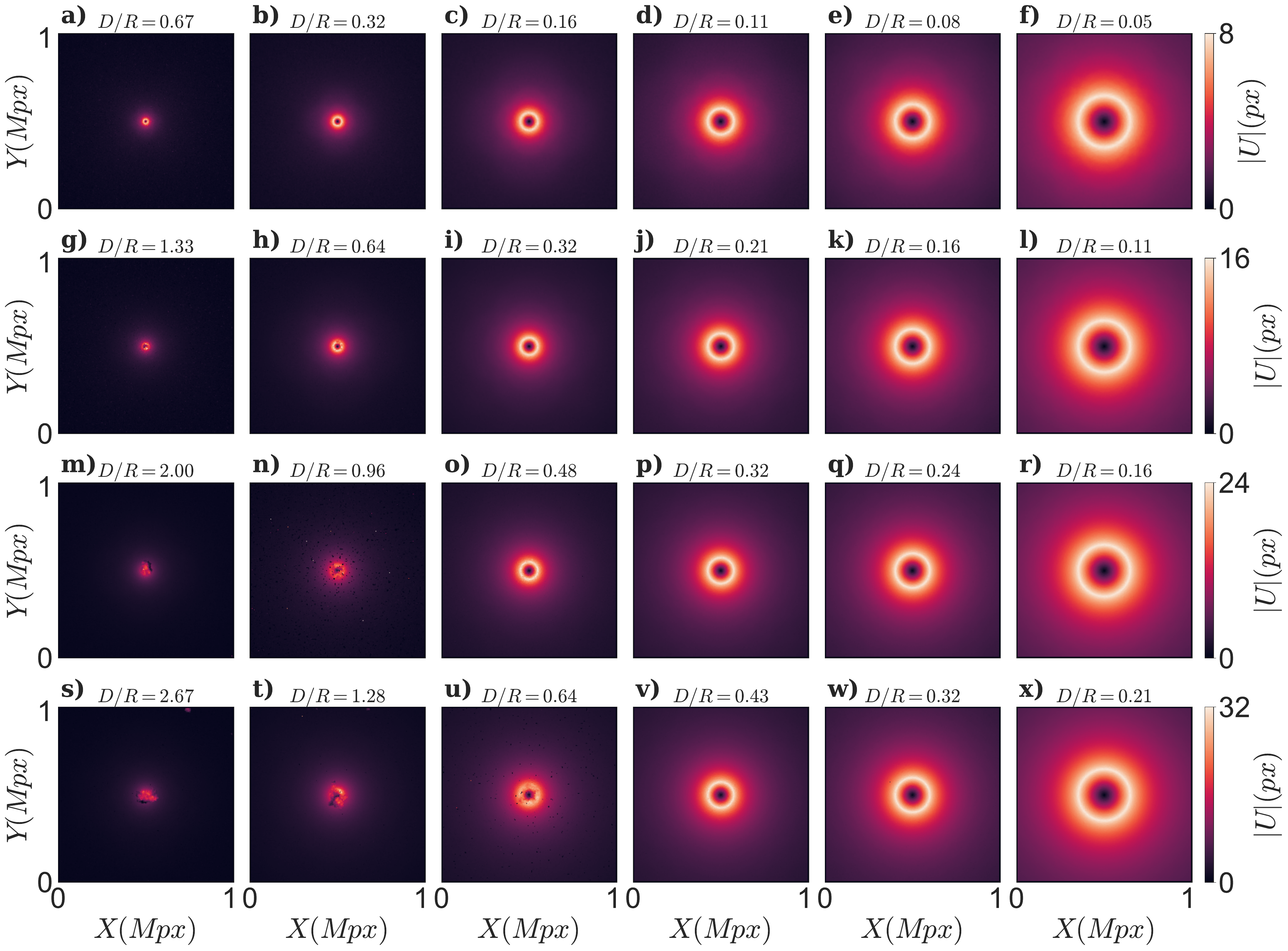}
    \caption{Best OF-PIV results for each case. The core radius of the vortex changes along the X axis (from $r = 12$ to $r=150$ pixels) and the maximum displacement on the field changes along the Y axis (from 8 to 32 pixels). \textbf{D/R} stands for the ratio between the maximum displacement and the Rankine core radius. }
    \label{fig:best_results}
\end{figure}

To go further, configurations minimizing the $\Psi$ error criteria were sought. Table~\ref{tab:of_best} presents the set of parameters leading to the best possible OF-PIV results for the present study. Fig.~\ref{fig:best_results} shows the contour of the velocity amplitude associated with all the parameters presented in Table~\ref{tab:of_best}. 
It can be seen that the majority of results come from images with a higher particle concentration, as was highlighted previously. For cases with concentrations below 15 $p/IW$, although Fig.~\ref{fig:best_results} and Tab.~\ref{tab:of_best} present the best results for each image configuration, the code limits may have been exceeded. Indeed, there seems to be a direct link between the size of the maximum displacement $D$ and the size of the structure that can be resolved (in this case, the radius $R$ of the vortex). In order to quantify this effect, the ratio between the maximum displacement and the radius of the vortex core $D/R$ is introduced. It can be considered as an indicator of the displacement gradient. Larger $D/R$ values correspond to small, rapidly rotating vortices while lower $D/R$ values correspond to large, slowly rotating vortices. The interaction between these two parameters can be clearly seen in Fig.~\ref{fig:best_results}, where there appears to be a limit close to $D/R \approx 0.7$ (sub-Fig.~\ref{fig:best_results} g,m,n,s and t) of the resolution capacity of the OF-PIV algorithm. As expected, the smaller, faster vortices are the most difficult to measure. \\

Another interesting observation that can be made from both Table~\ref{tab:of_best} and Fig.~\ref{fig:best_results} is that the choice of \textbf{KR} does not seem to be directly linked to the size of the displacement, as previously thought, but more to the \textit{gradient} of the displacement. This is very well illustrated in the case of the vortex with $r=100~px$ and a displacement magnitude of $32~px$, where the \textbf{KR} is only $2~ px$. However, if the results are ranked according to the $D/R$ ratio, it can be seen that there is another hidden trend pointing towards higher $D/R$ and smaller \textbf{KR} and vice versa. Since $D/R$ can be interpreted as a proxy for the displacement gradient, this means that the larger the displacement gradient, the smaller the kernel.
The latter can be explained by the nature of the algorithm, since larger kernel sizes tend to have a smoothing effect on OF results, a smaller size will help better detect very localized displacement variations, i.e. high gradients. This trend can be observed in Fig.~\ref{fig:best_results} b, i, p, w) which all have $D/R = 0.32$ and very similar OF parameters, regardless of size of the displacement or the radius of the vortex core.\\

\begin{figure}[!h]
    \centering
    \includegraphics[width = 0.9\textwidth]{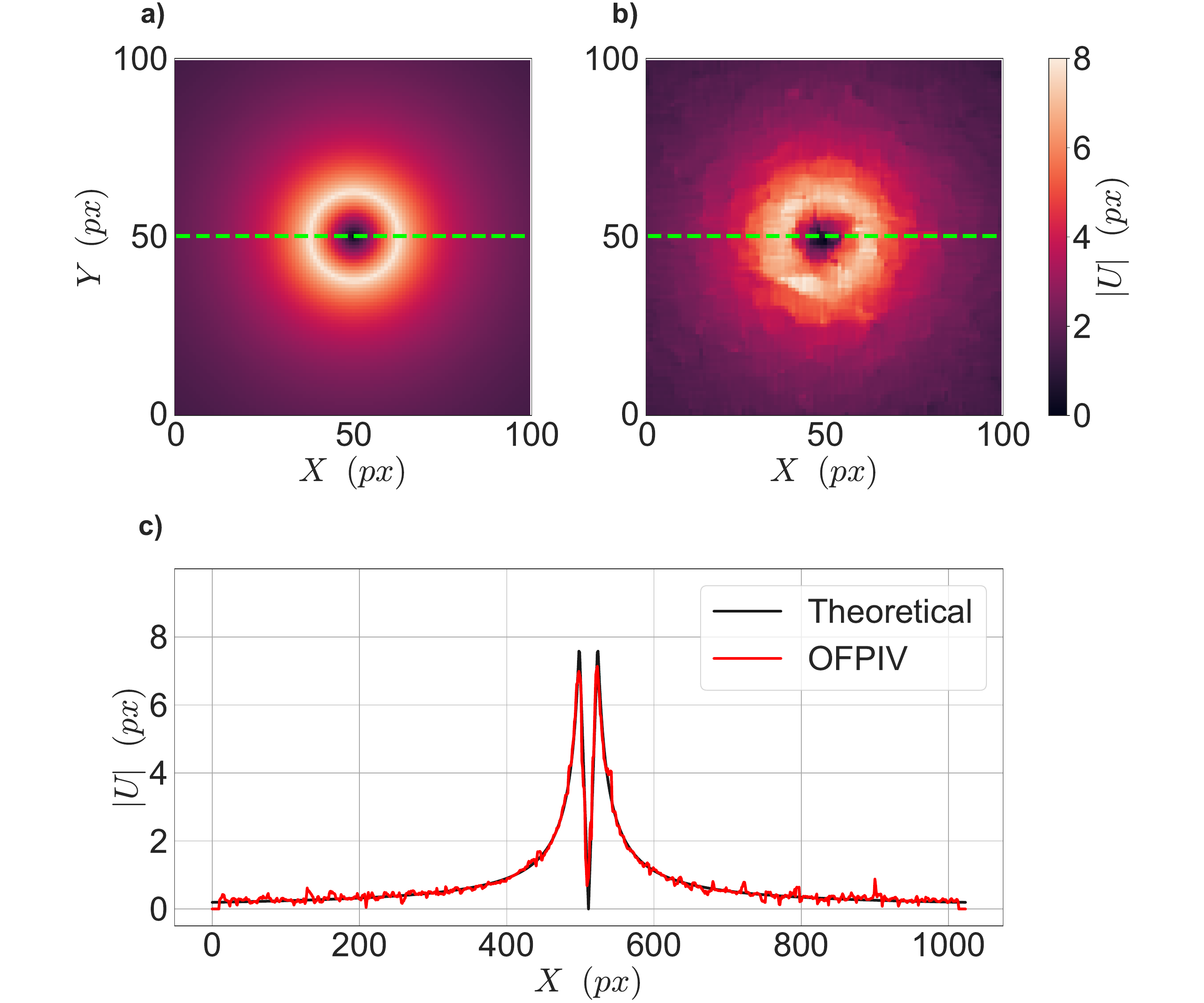}
    \caption{Comparison between theoretical displacement and OF-PIV results for a small Rankine vortex of core radius $r = 12$ pixels and a maximum expected displacement of 8 pixels. a) Theoretical displacement field zoomed-in around the vortex b) OF-PIV result zoomed-in around the vortex. c) Displacement magnitude profiles comparison over one line of pixels across the middle of the image height (dotted green line). OF-PIV result achieved from an image pair with a particle concentration of 15 $p/IW$}
    \label{fig:rk_comparison_12}
\end{figure}

\subsection{Resolution of small scales}

\begin{figure}[!h]
    \centering
    \includegraphics[width=0.9\textwidth]{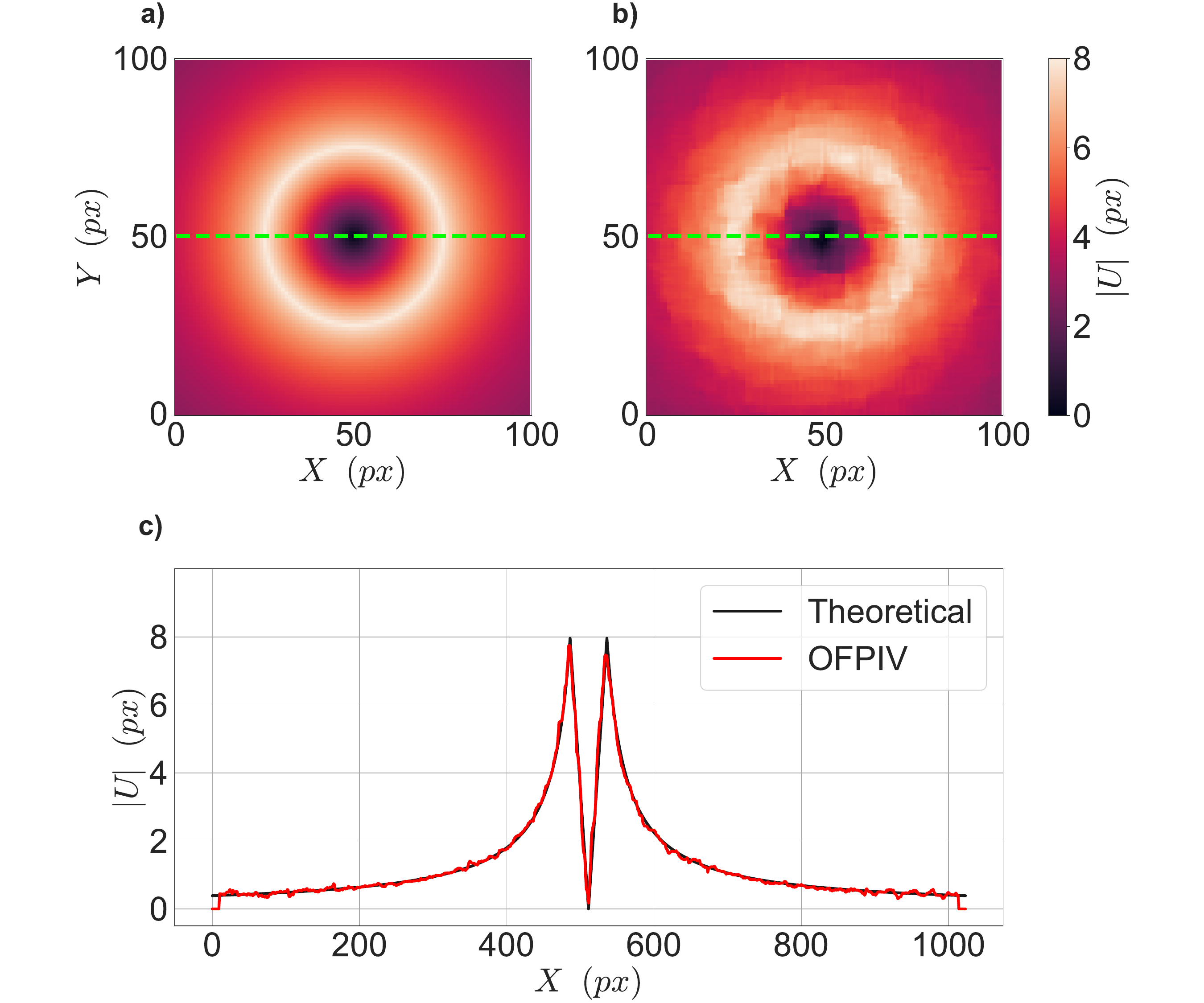}
    \caption{Comparison between theoretical displacement and OFPIV result for a Rankine vortex of core radius $r = 25$ pixels and a maximum expected displacement of 8 pixels. a) Theoretical displacement field zoomed-in around the vortex b) OF-PIV result zoomed-in around the vortex. c) Displacement magnitude profile comparison over one line of pixels across the middle of the image height (dotted green line). OF-PIV result achieved from an image pair with a particle concentration of 15 $p/IW$}
    \label{fig:rk_comparison_25}
\end{figure}

An important question is the ability of OF-PIV to accurately measure small structures with relatively large displacements. Fig.~\ref{fig:rk_comparison_12} and Fig.~\ref{fig:rk_comparison_25} present a comparison of the best result obtained for two Rankine vortices, one of core radius $r = 12$ and the other with core radius $r=25$ pixels, and a maximum displacement of 8 pixels for both. Both cases are clear examples of a locally concentrated displacement gradient. Both figures show a zoom-in on the theoretical vortex core and OF-PIV (sub-Figures a and b respectively). Sub-Figures c shows a comparison of a velocity profile extracted from a 1-pixel line halfway up the image (dashed green line).

In Fig.~\ref{fig:rk_comparison_12} and \ref{fig:rk_comparison_25} one can appreciate a very precise detection of the structures, taking into account their small sizes. This is particularly true for the $r=12$ pixel vortex shown in Fig.~\ref{fig:rk_comparison_12}. Even if the global maxima are not perfectly found (Fig.~\ref{fig:rk_comparison_12}c), the differences on the displacement peaks are within 5\% of the theoretical value. Detection of maxima positions and velocity gradient resolution are good, even when comparing velocity profiles on a single 1 pixel thick profile. The same could be said for the larger vortex ($r=25$ pixels) shown in Fig.~\ref{fig:rk_comparison_25}. OF-PIV makes it possible to better identify the structure (Fig.~\ref{fig:rk_comparison_25}b) and its corresponding global maxima and minima (Fig.~\ref{fig:rk_comparison_25}c). These two cases support the assertion of a limit for OF-PIV based on the gradient of the displacement, more than on its magnitude, since the only change between these two cases is the size of the structure.      \\

\begin{figure}[!h]
    \centering
    \includegraphics[width=1\textwidth]{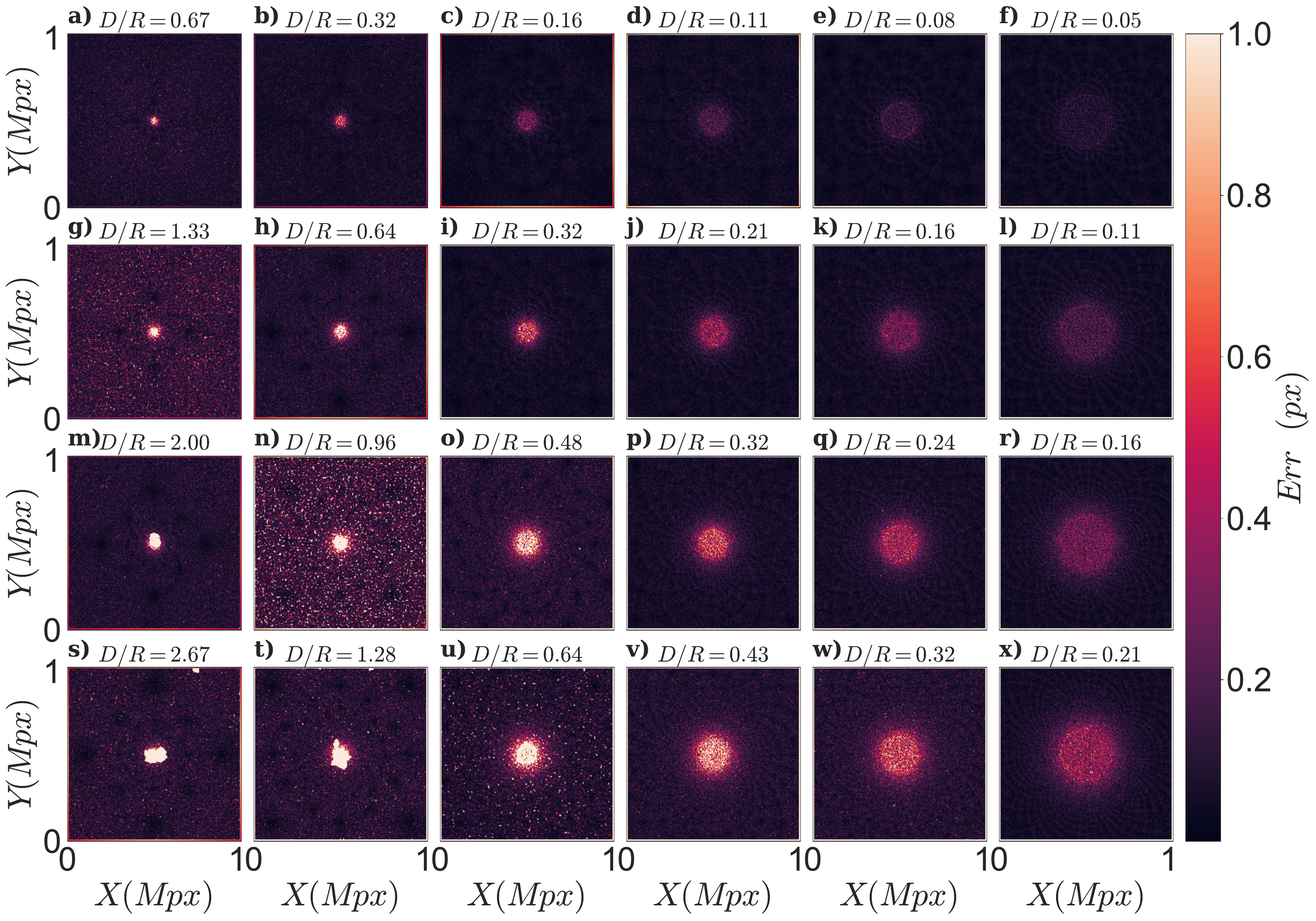}
    \caption{Absolute displacement error $Err$ of the best OFPIV results for each of the tested cases. The core radius of the vortex changes along the X axis (from $r = 12$ to $r=150$ pixels) and the maximum displacement on the field changes along the Y axis (from 8 to 32 pixels). \textbf{D/R} stands for the ratio between the maximum displacement and the Rankine core radius.}
    \label{fig:best_results_error}
\end{figure}

Fig~\ref{fig:best_results_error} presents the absolute displacement error $Err$ obtained for the best OF-PIV results for each of the image configurations. Here, the influence of the gradient $D/R$ of the displacement becomes even more obvious since the error follows a decrease proportional to the decrease in the gradient towards the right of the Figure. Furthermore, it can be clearly observed that the error levels decrease with $D/R$. For example, for Fig~\ref{fig:best_results_error} b,i,p,w) which correspond to $D/R = 0.32$, one can see that the error behaves very similarly in localization and in magnitude. These results are also very encouraging since they show that OF-PIV can give results with error margins of the order of a sub-pixel, if good conditions are met, both in terms of choice of OF parameters -PIV as input image settings.    \\

\section{Influence of particle seeding on experimental measurements}\label{sec:exp_con}

Given the nature of the OF algorithm, it is important to study in depth the impact of particle seeding on the quality of the results. This question is even more relevant when performing RT-OFPIV measurements for hours, looking for low-frequency signatures in the fluctuation of scalar quantities derived from instantaneous velocity fields. Indeed, the question of seeding becomes crucial due to the sedimentation of particles over time. To avoid loss of information in the velocity fields over time, it becomes necessary to add particles into the closed-loop hydrodynamic channel. Unfortunately, a quantitative criterion is missing to know how many particles should be injected and when.

\begin{figure}[!h]
    \centering
    \includegraphics[width=0.8\textwidth]{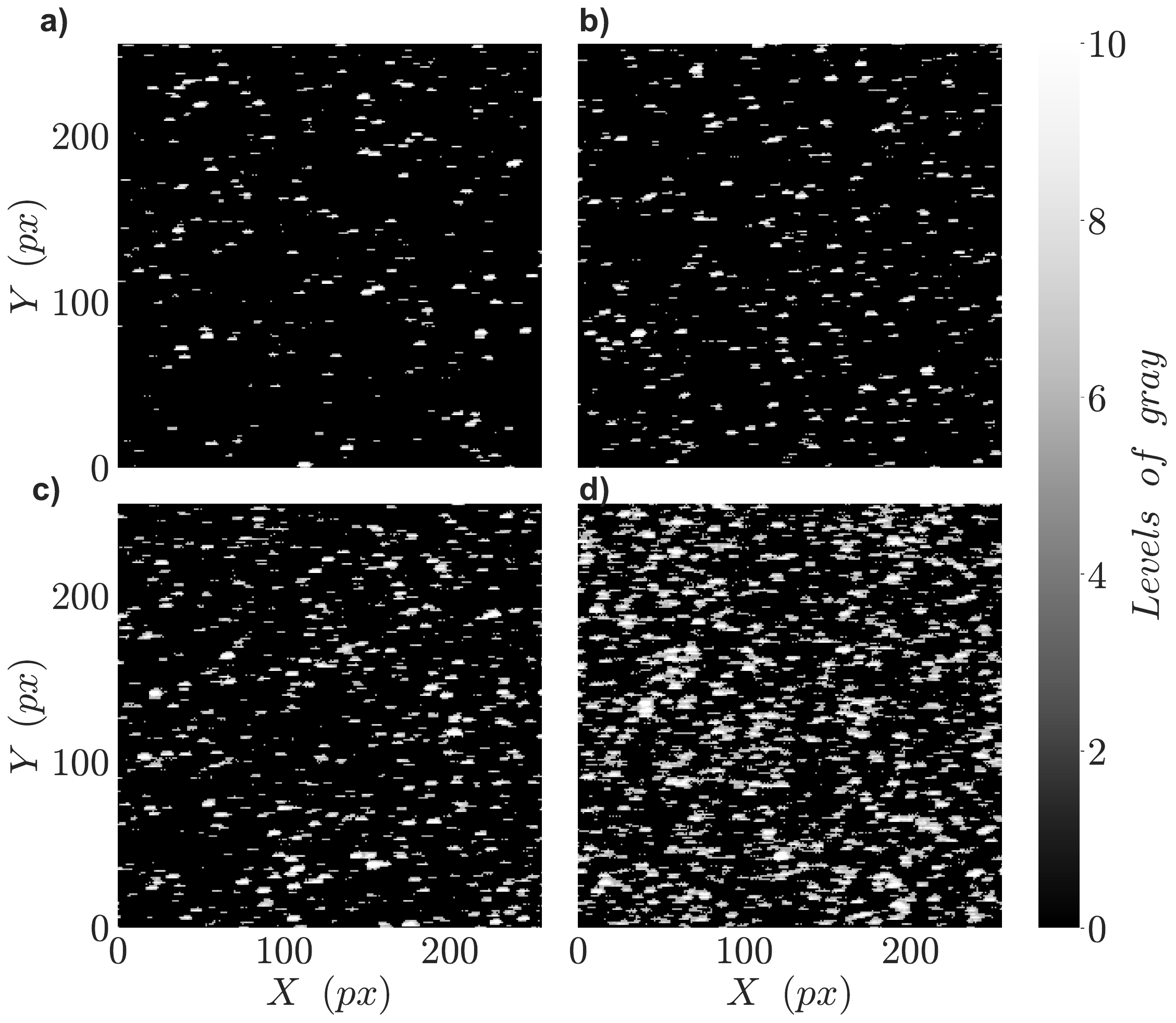}

    \caption{Sub-sample of the raw images during the change in particle concentration. a) $C_0$. b) $2C_0$. c) $3C_0$. d) $4C_0$ }
    \label{fig:raw_small}
\end{figure}

The first observation is that the standard criteria used to adjust the concentration of particles in CC-PIV (particles per window or particles per pixel) are not adapted to OF-PIV.    The appropriate criteria for OF-PIV should be related to the number of pixels \textit{containing information relating to intensity variations}. This notion must be taken into account because the OF algorithm works optimally when fed with a very textured image, that is to say when each pixel sees an intensity variation that can be linked to movement. Unfortunately, one can see that many pixels are black in the standard snapshots (Fig.~\ref{fig:raw_small}a) used with a concentration $C_0 $ adapted to CC-PIV. All black pixels \textit{do not provide information} to the OF algorithm and are useless for calculating instantaneous velocity fields. It becomes necessary to define new quantitative criteria to optimize particle seeding in order to obtain the best possible textured images, thus leading to better instantaneous velocity fields calculated with OF. The ultimate goal is for each pixel to contain relevant information, leading to a maximum spatial resolution of 1 vector per pixel.

The first step consisted of increasing the concentration of particles by successive injection of particles inside the volume of water. The particle concentration will be expressed as mass of particles per volume of water ($g/L$). The first injection ($C_0 = 9.21~10^{-3} g/L$) corresponds to the standard concentration used for PIV measurements. Then, the same amount of particles $C_0$ was successively injected every 20 $min$, while image pairs were captured every 5 $s$. The measurements were carried out in a horizontal plane, in the freestream, vortex-free region, upstream of the cylinder (Fig.~\ref{fig:Cyl}). One can see in Fig.~\ref{fig:raw_small} the evolution of the concentration in small windows (256$\times$256 pixels$^2$) taken at the center of the raw snapshots. As expected, as the concentration increases, the number of black pixels decreases.

In order to quantify the number of pixels containing information, the background noise of the camera sensor is removed (4 gray levels with a dynamic range of 8 bits). Then, the number of active pixels $N_{act}$ that actually detect intensity changes above the noise level is calculated. The ratio $R_{act} = N_{act} / N_{pix}$ of the number of active pixels (i.e. with information) to the total number of pixels $N_{pix}$ of the sensor gives the percentage of camera pixels actually containing information useful to the OF.

\begin{figure}[h]
    \centering
    \includegraphics[width=0.9\textwidth]{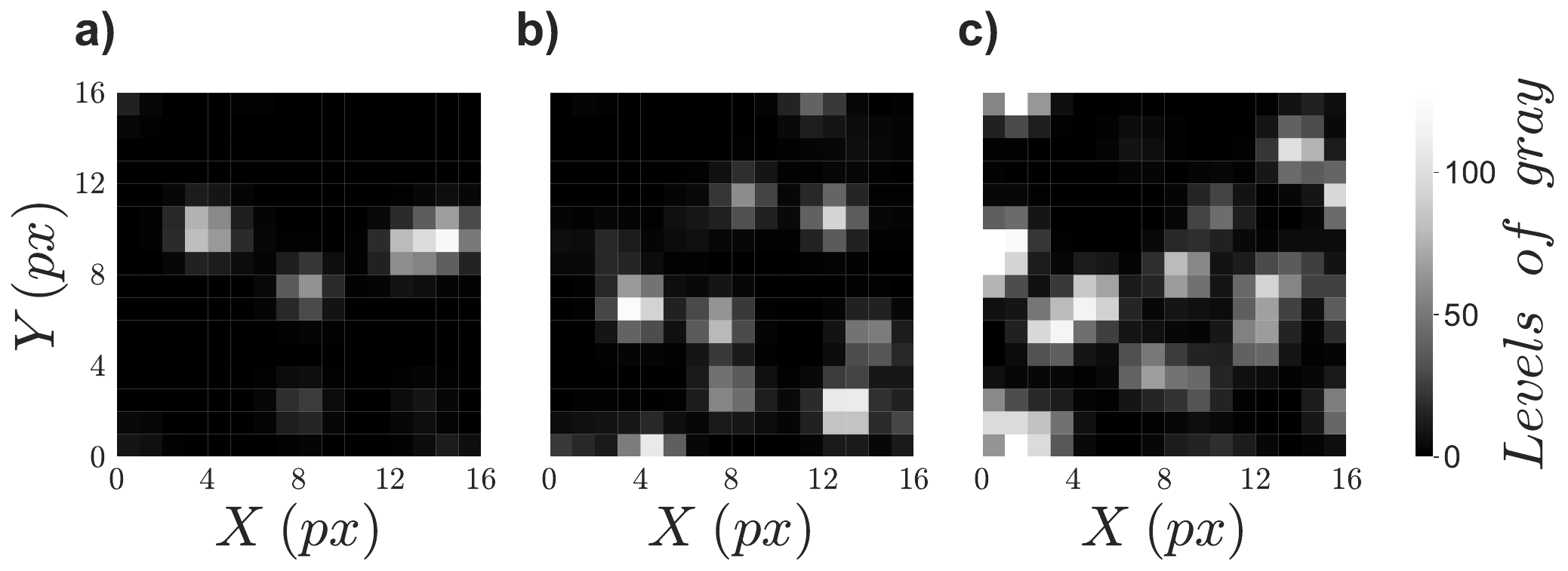}
    \caption{Examples of raw snapshots obtained with the synthetic image generator for increasing particle density: 5 p/IW (a), 10 p/IW (b) and 15 p/IW (c). It corresponds respectively to $R_{act} = 36 \%$, $R_{act} = 60 \%$ and $R_{act} = 73 \%$. }
    \label{fig:window_comparison}
\end{figure}

First, $R_{act}$ is calculated for the synthetic PIV images used in the previous section. Fig~\ref{fig:window_comparison} shows examples of particle distribution in 16$\times$16 pixels$^2$ interrogation windows obtained for the three concentrations used to create the images. It can be seen that to increase the particle concentration, from 5 $p/IW$ to 15 $p/IW$, the number of active pixels is also increased, from $R_{act} = 36 \%$ to $R_{ act} = 73\%$.

\begin{figure}[!h]
    \centering
    \includegraphics[width=1\textwidth]{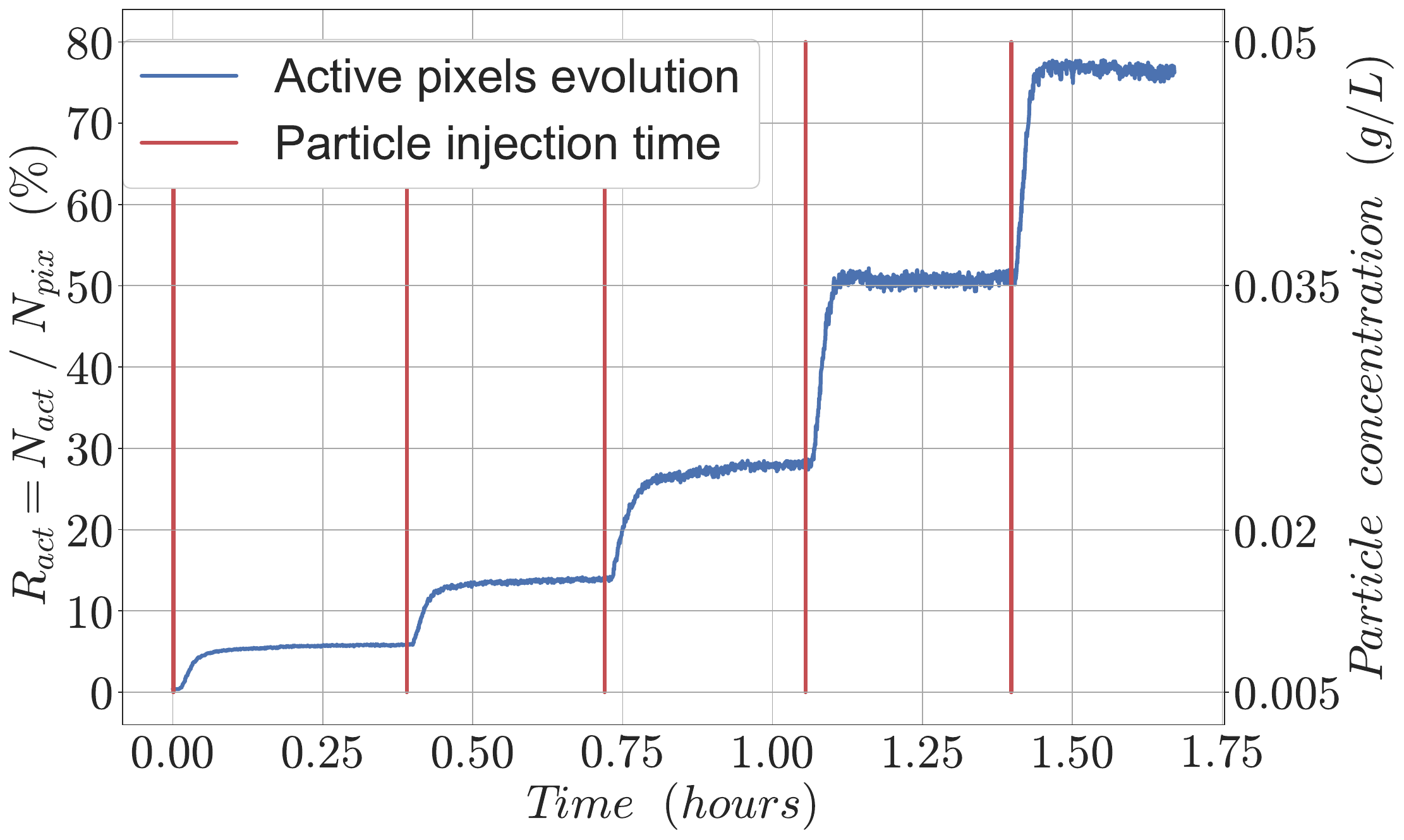}

    \caption{ Evolution of percentage of active pixels $R_{act}$ for increasing concentration of particles. Each red vertical line indicate the time of a new injection of particles in the water tunnel, leading to an increase of the particle concentration.}
    \label{fig:particle_density_evolution}
\end{figure}

Fig.~\ref{fig:particle_density_evolution} shows the evolution of $R_{act}$ as a function of time to increase the particle concentration. Each red vertical line corresponds to a new injection of particles inside the hydrodynamic channel, leading to an increase in the concentration of particles seen by the camera sensor. The first observation is that for the initial concentration, commonly used for CC-PIV, only 6\% of the sensor pixels contain information. This is clearly insufficient for an OF algorithm to calculate 1 vector per pixel. One can see that after 5 injections, leading to a total particle concentration of around 0.05 $g/L$ (10 $\times C_0$), the ratio of active pixels is greatly increased to almost reach $R_{act } =$80\%.

\begin{figure}[!h]
    \centering
    \includegraphics[width=1\textwidth]{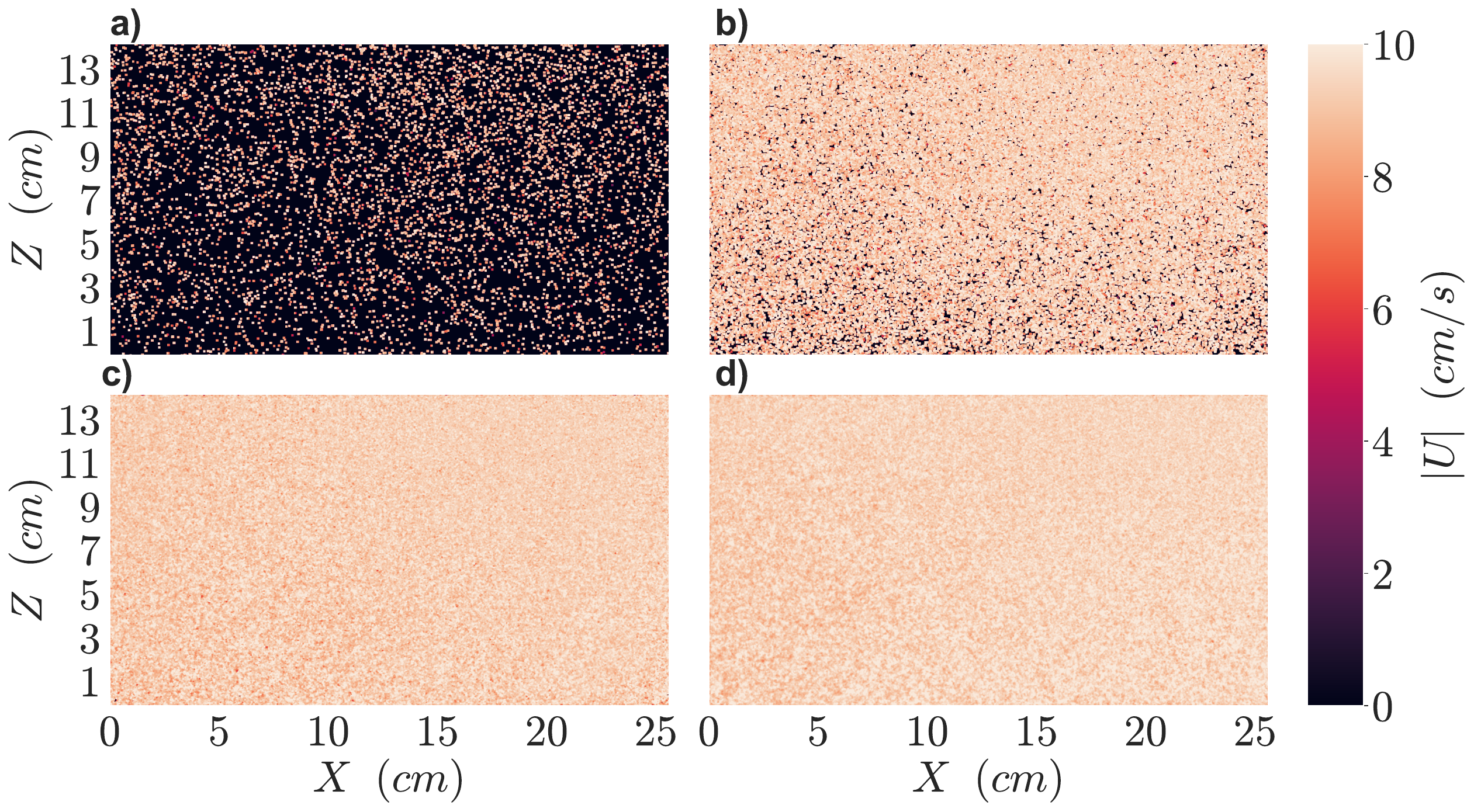}

    \caption{Impact of the number of active pixels on the velocity fields in the free-stream region at different times for increasing concentration of particles. a) $C_0$ b) 2 $C_0$ c) 7 $C_0$ d) 10 $C_0$.}
    \label{fig:density_quality_difference}
\end{figure}

This strong evolution in the proportion of active pixels should impact the quality of the resulting instantaneous velocity fields calculated with the OF algorithm. Fig. ~\ref{fig:density_quality_difference} shows the effect of particle injection on the resulting velocity fields in the free flow region. It can be seen that the velocity field becomes denser and more homogeneous as the particle concentration increases. For the maximum concentration, the velocity field becomes much smoother and uniform. No smoothing was applied to the instantaneous velocity fields.

As stated previously, the Kernel radius is a key parameter for OF. When dealing with suboptimal conditions for image texture (low particle concentration), a larger kernel radius is needed to resolve the velocity estimate in a pair of images on each pixel. This has the side effect of losing smaller structures, because it smoothes the flow field. Using the appropriate concentration, one should obtain valuable information close to pixel resolution using a smaller kernel radius.

\section{Influence of the concentration of particle on the spatial resolution in the velocity fields}

For both upcoming subsections the background camera noise was suppressed, as in the previous section, to enhance the observation of the particle seeding impact.

\begin{figure}[!h]
    \centering
    \includegraphics[width=1\textwidth]{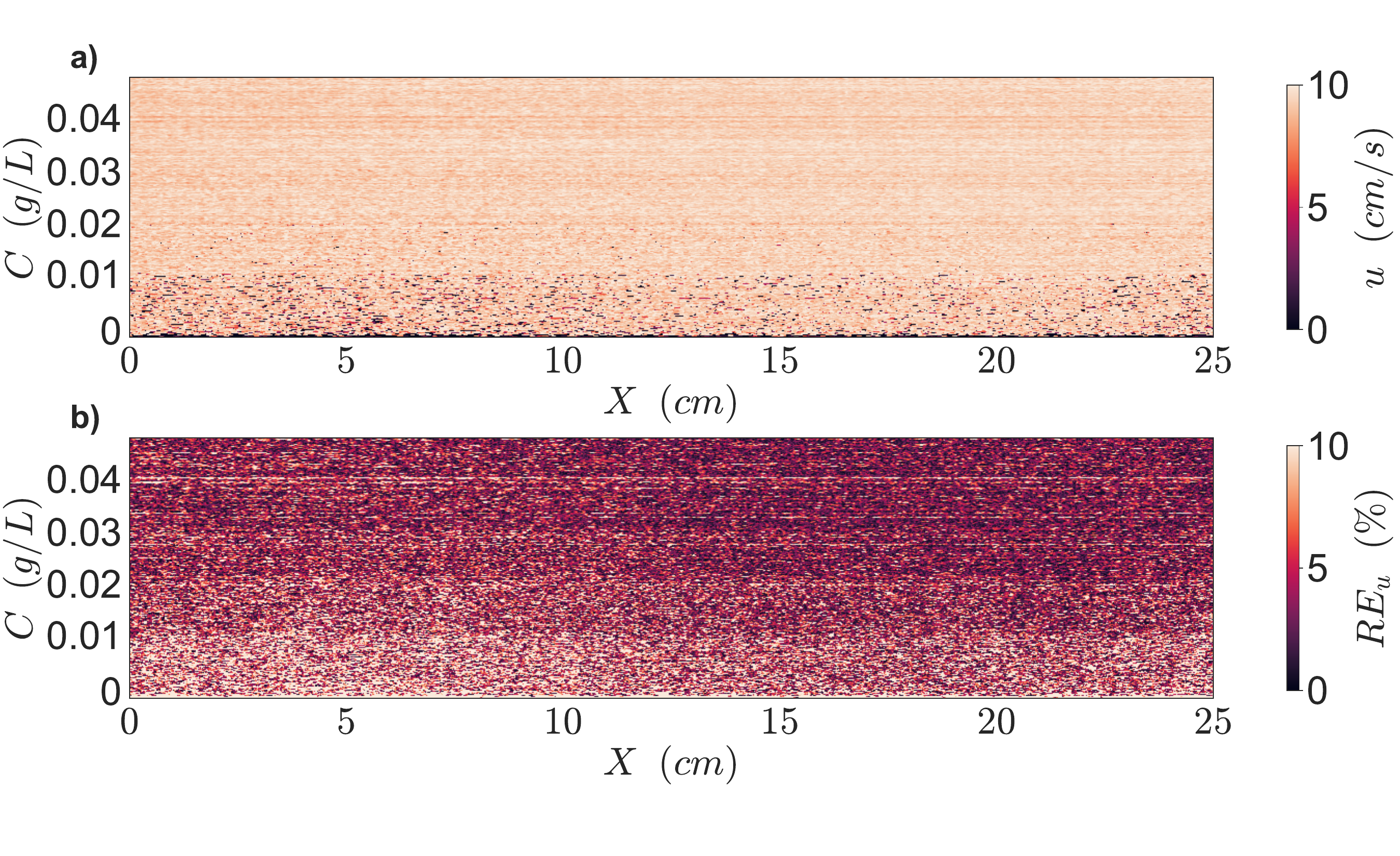}

    \caption{a) Longitudinal velocity profile at the center of the image for increasing concentration of particles. b) Corresponding variation of relative error of the velocity profile at the image center for increasing concentration of particles. }
    \label{fig:scanline_5-25}
\end{figure}

\subsection{Freestream flow}
In order to evaluate the impact of seeding in terms of quality of the resulting velocity fields, images in the free zone of the tunnel, upstream of the cylinder, were taken. The objective was to calculate the velocity fields of a homogeneous flow, in a region where there are no complex velocity gradients or other 3D phenomena difficult to estimate. These are the same images that were taken to measure the amount of active pixels in the images in the previous section. 

Fig.~\ref{fig:scanline_5-25}a) presents a 2D plot of the evolution of the instantaneous velocity profile in the direction of the current, extracted from a line of pixels in the center of the image, to increase the concentration of particles. One can see that a minimum concentration is necessary to homogenize the speed profiles and obtain a good estimate of the speed. This is confirmed by Fig.~\ref{fig:scanline_5-25}b) which shows the evolution of the relative error $(RE_u = \frac{|U_\infty - u|}{U\infty} * 100)$ made on the estimation of the speed in the direction of the current. The error is minimized for the largest particle concentration. 

A more in-depth analysis, taking into account OF parameters, is still necessary. Nevertheless, it is clear that for such uniform flow, increasing particle seeding in the water tunnel leads to a better resolved velocity field and reduced noise and errors.

\subsection{Flow past a cylinder}

The same measurements were carried out downstream of a cylinder in a horizontal plane at mid-height of the cylinder (Fig.~\ref{fig:Cyl}), in order to estimate the influence of the seeding concentration on the  instantaneous velocity fields quality. The objective was to compare the spatial resolution of the velocity fields for a flow containing numerous eddies of various scales. The measurements were carried out at Reynolds $Re_h = \frac{D \times U_\infty}{\nu} = 1443$, which corresponds to the subcritical regime, but with high velocity fluctuations at different scales. The images have a size of $X =5120,~Y=1440$ pixels, which results in images of $7.37~Mpx$.

\begin{figure}[!h]
    \centering
    \includegraphics[width=1\textwidth]{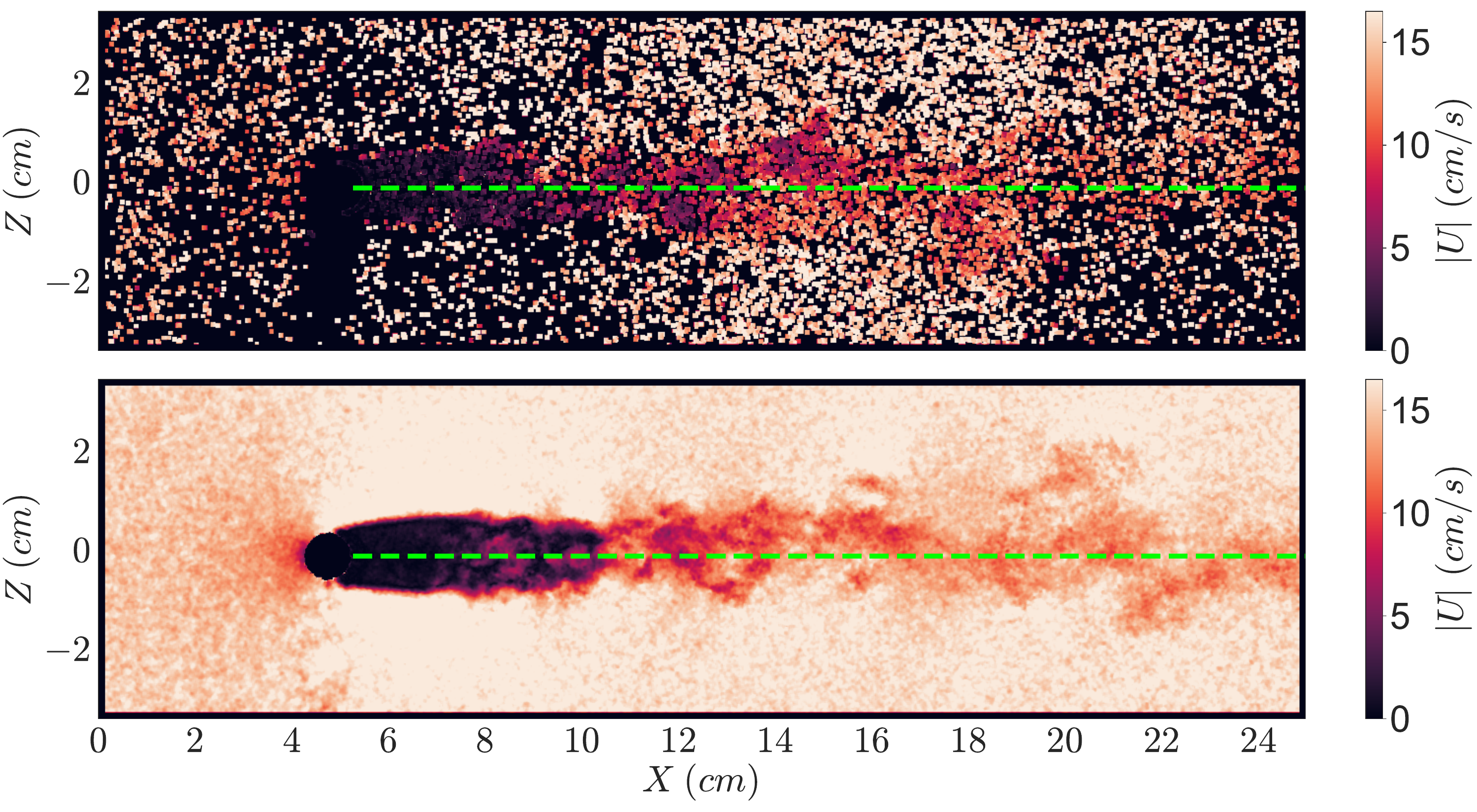}

    \caption{2D plots of the instantaneous velocity magnitude field  obtained with a low ($C_0)$  concentration (a) and for a large ($5C_0$) concentration (b). The qualitative difference is obvious if no smoothing is applied on the velocity fields. With a larger concentration, the instantaneous velocity field shows no hole and one can distinguish very fine structures and small velocity variations. Movie online.}
    \label{fig:Cyl_mag}
\end{figure}

Two configurations are compared: one with the standard particle concentration ($C_0$), the other with a higher concentration ($5C_0$).

An instantaneous velocity magnitude field obtained with a low concentration is shown in Fig.~\ref{fig:Cyl_mag}a). It can be seen that if no smoothing is applied, some information is lost, leading to holes or errors in the velocity field.  It is very different as the concentration increases: there are no holes in the field which appears smooth and has fine details.

\begin{figure}[!h]
    \centering
    \includegraphics[width=0.9\textwidth]{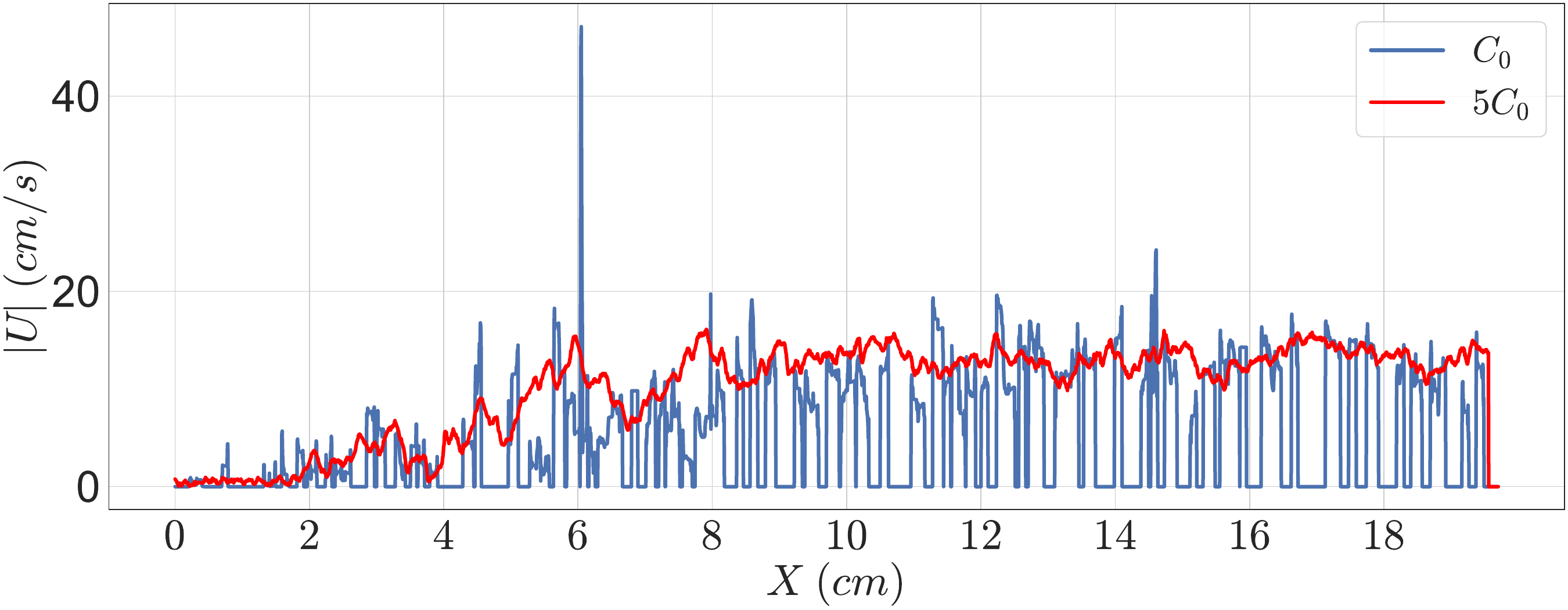}

    \caption{Instantaneous profiles of the velocity magnitude measured along a pixel line at $z=D/2$. In blue the low concentration ($C_0$) and in red the high concentration ($5C_0$). One can see that the red profile is both smoother, without holes or sharp nonphysical variations when the seeding is increased. Movie online.}
    \label{fig:cyl_prof}
\end{figure}

The velocity magnitude profile along the center line downstream of the cylinder (green line in Fig.~\ref{fig:Cyl_mag}) is plotted for low and high concentrations (Fig.~\ref{fig:cyl_prof}). This confirms that when the concentration is too low, many pixels do not contain information which leads to many zero pixels. On the other hand, it is clear that the resolution of the velocity field is improved with higher concentration. This leads to smoother velocity magnitude profiles that clearly contain interesting velocity fluctuations, without any smoothing or interpolation. Here, the resulting spatial resolution for the velocity vectors is 47.62 $~\mu m$, or 400 vectors per $mm^2$. It is important to note that both cases were treated with the same OF parameters (\textit{Normalization radius} of 3 pixels, \textit{Kernel radius} of 6 pixels, 4 \textit{Pyramid sublevels} and 3 \textit{ iterations}).

\section{Conclusion}

The objectives of the present study were to evaluate the quality of OF-PIV results through a benchmark with synthetic images and to optimize the experimental conditions to adapt the requirements of OF-PIV. More specifically, we were interested in the influence of images and OF-PIV parameters on the results, as well as the influence of particle seeding on the spatial resolution. 

First, an error test was carried out to evaluate and quantify the quality of the OF-PIV results. The study was carried out using synthetically generated PIV images of a Rankine vortex. The case of the Rankine vortex was analyzed according to different displacement magnitudes and vortex core sizes. The concentration of particles in the images was found to be of primary importance in improving the OF-PIV results. A trend leading to a possible association of the size of the Kernel radius with the size of the displacement gradient was found. OF-PIV results can be within a sub-pixel margin of error when the parameters of the algorithm and input images are chosen correctly.

The relevance of particle seeding density on the quality and resolution of OF-PIV was also studied. For this purpose, the notion of \textit{active pixels} was introduced as a proxy for image quality. The objective was to search for a particle seeding criterion different from the criteria used for CC-PIV, adapted to OF. It has been shown that it is indeed possible to increase the number of actually useful active pixels in a given pair of images. Using standard seeding, well suited to CC-PIV, less than 10\% of the camera sensor was used. Increasing the particle concentration led to more than 80\% active pixels. 

Finally, it was shown that thanks to this optimization, it was possible to increase the spatial resolution leading to much better instantaneous velocity fields. Measurements of the flow past a cylinder were used as a reference to evaluate the quality of the instantaneous velocity fields. 2D velocity fields obtained with lower concentration exhibit holes and errors that disappear if the particle concentration increases. Additionally, fine details associated with small structures can be observed, showing that with the appropriate parameters, OF-PIV measurements can effectively lead to dense velocity fields, with 1 vector per pixel. More systematic experiments are still needed to generalize these results. For example, this study demonstrates that the small scales of turbulent flow can indeed be resolved in RT-OFPIV measurements, but these experiments could not be performed in the present study.

\section{Acknowledgements}
This study benefits from the support of the ANRT (french National Agency for Technological Research) which co-finances the Cifre thesis of J. Pimienta. The research also benefits from an industrial collaboration with Photon Lines Inc.
\bibliography{bibliography}
\end{document}